\documentclass[12pt]{article}

\usepackage{amsmath,amssymb}
\usepackage{graphicx}
\usepackage{slashed}
\usepackage[margin=2.25cm]{geometry}
\usepackage[blocks]{authblk}
\usepackage{caption}
\usepackage{subcaption}
\usepackage[compat=1.1.0]{tikz-feynman}
\usepackage{cite}
\usepackage{xcolor}

\date{\today}

\newcommand{\ie}{\textit{i.e.,}}
\newcommand{\eg}{\textit{e.g.,}}
\newcommand{\dt}{\ensuremath{\mathrm{d}t}}
\renewcommand{\Im}{\mathrm{Im}}

\newcommand{\lsr}{\ensuremath{\mathcal{R}}}
\newcommand{\gsr}{\ensuremath{\mathcal{G}}}
\newcommand{\fesr}{\ensuremath{\mathcal{F}}}
\newcommand{\double}[2]{\ensuremath{(#1,\,#2)}}
\newcommand{\triple}[3]{\ensuremath{(#1,\,#2,\,#3)}}
\newcommand{\ee}{\ensuremath{\mathrm{e}}}
\newcommand{\ii}{\ensuremath{\mathrm{i}}}
\newcommand{\msbar}{\ensuremath{\overline{\text{MS}}}}
\newcommand{\gev}{\ensuremath{\text{GeV}}}
\newcommand{\flav}{\ensuremath{d u\bar d\bar u}}
\newcommand{\tq}{\ensuremath{T^{0^{+-}}_{\flav}}}

\begin{document}

\title{A Sum-Rules Analysis of
Next-to-Leading-Order (NLO) QCD Perturbative Contributions to a  
$J^{PC}=0^{+-}$, $\flav$ Tetraquark Correlator}
\author{K.~Ray\thanks{dkr504@mail.usask.ca}}
\author{D.~Harnett\thanks{derek.harnett@shaw.ca}}
\author{T.G.~Steele\thanks{tom.steele@usask.ca}}
\affil{Department of Physics and Engineering Physics, University of Saskatchewan, Saskatoon, SK,
S7N~5E2, Canada}

\maketitle

\begin{abstract}
We calculated next-to-leading-order (NLO) QCD perturbative contributions to a 
$J^{PC}=0^{+-}$, $\flav$ tetraquark (diquark-antidiquark) correlator 
in the chiral limit of massless $u$ and $d$  quarks.
At NLO, there are four quark self-energy diagrams and six gluon-exchange diagrams.
Nonlocal divergences were cancelled using diagrammatic renormalization.
Dimensionally regularized integrals were numerically computed using pySecDec.
The combination of pySecDec with diagrammatic renormalization establishes a 
valuable new methodology for NLO calculations of QCD correlation functions.
Compared to leading-order (LO) perturbation theory, we found that NLO perturbation 
theory is significant.
To quantify the impact of NLO perturbation theory on physical predictions, 
we computed NLO perturbative contributions to QCD Laplace, Gaussian, and finite-energy sum rules. 
Using QCD sum rules, we determined upper and lower bounds on the $0^{+-}$, $\flav$ 
tetraquark ground-state mass, $M$:
at NLO in perturbation theory, we found $2.2~\gev\lesssim M\leq 4.2~\gev$
whereas, at LO, we found $2.4~\gev\lesssim M\leq 4.6~\gev$.
This mass range suggests the possibility of mixing between $0^{+-}$,
light-quark (\ie\ $u$ and $d$ quarks) 
hybrid and $\flav$ tetraquark states.
Taking into account uncertainties in QCD parameters,
we found no evidence for a $0^{+-}$, $\flav$ tetraquark under 1.9~GeV.

\end{abstract}

\section{Introduction}\label{introduction}

Colour confinement allows for hadron families beyond the well-known 
two-quark mesons and three-quark baryons~\cite{Gell-Mann:1964ewy}.
Four-quark states are one such family (see \eg \cite{Jaffe:1976ig,Jaffe:1976ih}).
Experimental evidence for the existence of four-quark states is strong (see, for example, the 
reviews~\cite{Chen:2016qju,Lebed:2016hpi,Ali:2017jda,Olsen:2017bmm,Liu:2019zoy,Guo:2017jvc,Brambilla:2019esw}). 
Several states containing (a minimum of) four quarks have been reported including 
the $Z_c(3900)^{+}$ with quark content
$cu\bar{c}\bar{d}$~\cite{BESIII:2013ris,Belle:2013yex},
the $X(5568)^{+}$ with quark content $su\bar{b}\bar{d}$~\cite{D0:2016mwd,D0:2017qqm},
and the $X(6900)$ with quark content $cc\bar{c}\bar{c}$~\cite{LHCb:2020bwg}.

While all known manifestly four-quark states contain at least one heavy quark, 
a four-quark framework (with an inverted mass hierarchy for the scalar mesons) 
is also expected in systems without heavy quarks~\cite{Jaffe:1976ig,Maiani:2004uc}.
However, clearly identifying four-quark states that do not contain heavy quarks
is difficult due in part to overlapping hadron multiplets and, presumably, hadron mixing.
Given these difficulties, a promising search strategy is to look for bosonic hadrons
with exotic quantum numbers, \ie\ $J^{PC}$ combinations such as $0^{--}$, $0^{+-}$, and $1^{-+}$
that are forbidden for two-quark mesons. While not guaranteed to be four-quark states 
(\eg\ hybrid mesons can have exotic
$J^{PC}$~\cite{Lebed:2016hpi,Olsen:2017bmm,Brambilla:2019esw}),
hadrons with exotic quantum numbers are at least guaranteed to not be two-quark mesons.

One possible picture for the structure of a four-quark state 
is that of a diquark-antidiquark bound state, \ie\ a tetraquark~\cite{Jaffe:1976ig,Jaffe:1976ih,Maiani:2004uc}.
Tetraquarks with exotic quantum numbers that do not contain heavy quarks 
have been studied using QCD sum 
rules~\cite{Chen:2008qw,Chen:2008ne,Jiao:2009ra,Du:2012pn,Huang:2016rro,Fu:2018ngx}\footnote{They have also been studied with the MIT bag model~\cite{Aerts:1979hn} and a Coloumb gauge QCD Hamiltonian model~\cite{Cotanch:2006wv,General:2007bk}.}.
QCD sum rules are transformed dispersion relations that relate hadron properties
to QCD correlation functions of interpolating
currents~\cite{Shifman:1978bx,Shifman:1978by,Reinders:1984sr,Bertlmann:1984ih,Narison:2002woh,Gubler:2018ctz}.
Variants include Laplace, Gaussian, and finite-energy sum rules.
In~\cite{Chen:2008qw}, an analysis of light-quark (\ie\ $u$ and $d$ quarks)
and hidden-strange, isovector tetraquarks with $J^{PC}=1^{-+}$
yielded mass predictions of about 1.6~GeV for $qq\bar{q}\bar{q}$ states and 
2.0~GeV for $qs\bar{q}\bar{s}$ states where $q$ represents $u$ or $d$.
However, for many of the currents considered, the QCD spectral functions (\ie\ imaginary parts)
of corresponding correlators were negative (\ie\ unphysical) in the squared-energy scale, $t$, range 
$1~\text{GeV}^2\lesssim t\lesssim 4~\text{GeV}^2$.
For $t\gtrsim 4~\text{GeV}^2$, the QCD spectral functions were positive (and hence physical), 
but the hadron masses obtained were greater than 2.5~GeV.
As such, Ref.~\cite{Chen:2008qw} concluded that these currents did not provide 
evidence for the existence of $1^{-+}$ tetraquarks under 2~GeV. 
In~\cite{Chen:2008ne}, an analysis of light-quark and hidden-strange, 
isoscalar tetraquarks with $J^{PC}=1^{-+}$
yielded a $qs\bar{q}\bar{s}$ state mass prediction of 1.8~GeV--2.1~GeV.
However, similar to what was seen in~\cite{Chen:2008qw}, many currents led to correlators with 
negative QCD spectral functions for $2~\text{GeV}^2\lesssim t\lesssim 4~\text{GeV}^2$.
All of the $qq\bar{q}\bar{q}$ currents showed this unphysical behaviour, and,
consequently, in Ref.~\cite{Chen:2008ne}, no mass predictions were obtained for $1^{-+}$,
light-quark, isoscalar tetraquarks.
In~\cite{Jiao:2009ra}, light-quark and hidden-strange, isovector and isoscalar 
tetraquarks with $J^{PC}=0^{--}$ were studied 
using a set of scalar currents, but none of the sum-rules analyses stabilized.
In~\cite{Huang:2016rro}, $0^{--}$, light-quark tetraquark states were studied 
using a set of vector currents which yielded isospin-degenerate
mass predictions of $(1.66 \pm 0.14)$~GeV.
In~\cite{Du:2012pn}, light-quark tetraquarks (as well as hidden-charm and
hidden-bottom tetraquarks) with $J^{PC}=0^{+-}$ were studied using scalar currents
(that contained covariant derivative operators). 
For the $qq\bar{q}\bar{q}$ states, no sum-rules analyses were successful.
In~\cite{Fu:2018ngx}, light-quark and hidden-strange tetraquarks with $J^{PC}=0^{+-}$ 
were studied using a set of vector currents.
For $qq\bar{q}\bar{q}$ tetraquarks, a mass of $(1.43\pm0.09)$~GeV was reported, and,
for $qs\bar{q}\bar{s}$ tetraquarks, a mass of $(1.54\pm0.12)$~GeV was reported.

\begin{figure}[htb]
\centering
\begin{subfigure}{0.45\textwidth}
    \centering
    \includegraphics[width=\textwidth]{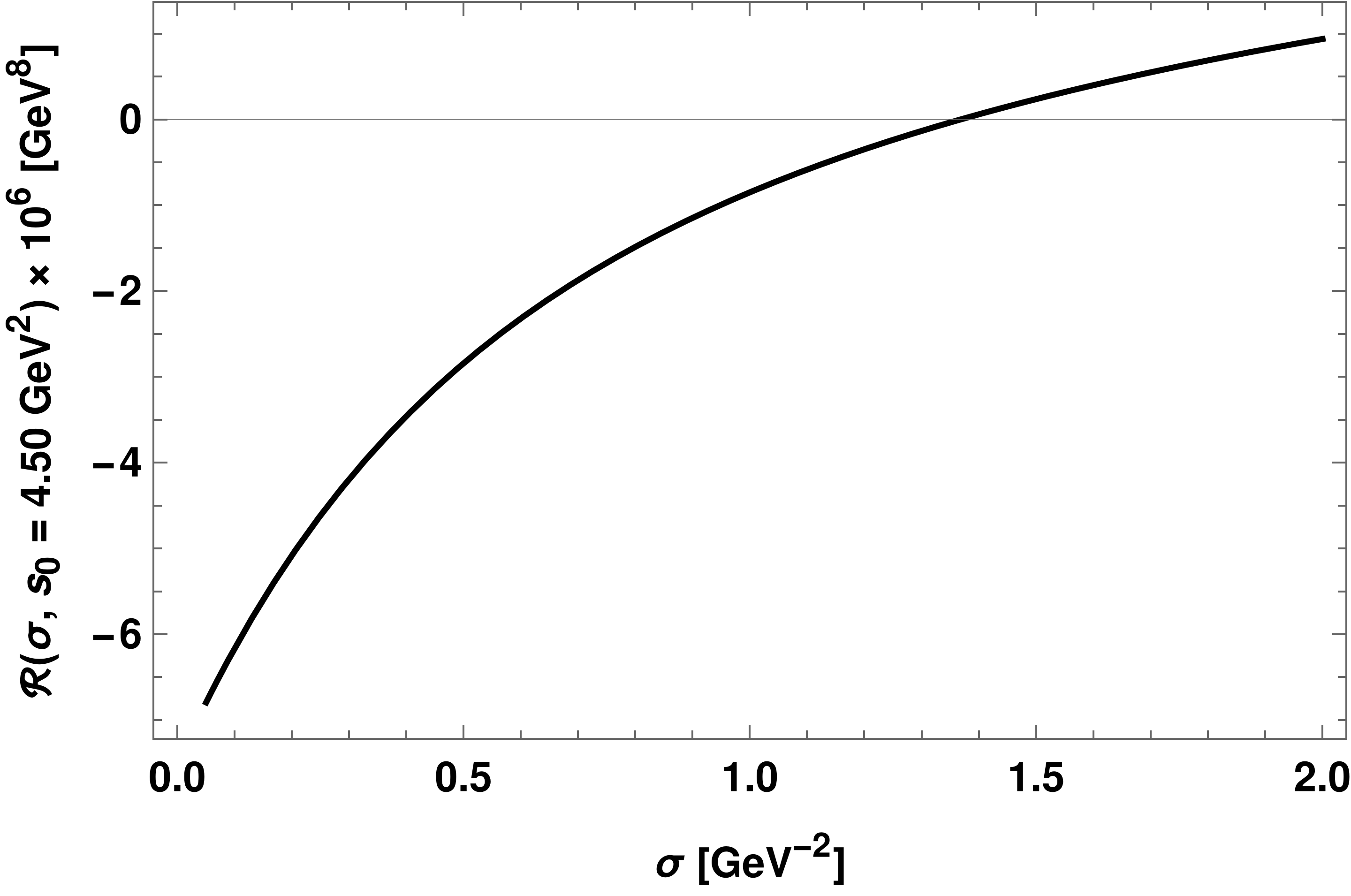}
    \caption{The order-0 (subtracted) LSR.}
    \label{faafo:a}
\end{subfigure}
\hfill
\begin{subfigure}{0.45\textwidth}
    \centering
    \includegraphics[width=\textwidth]{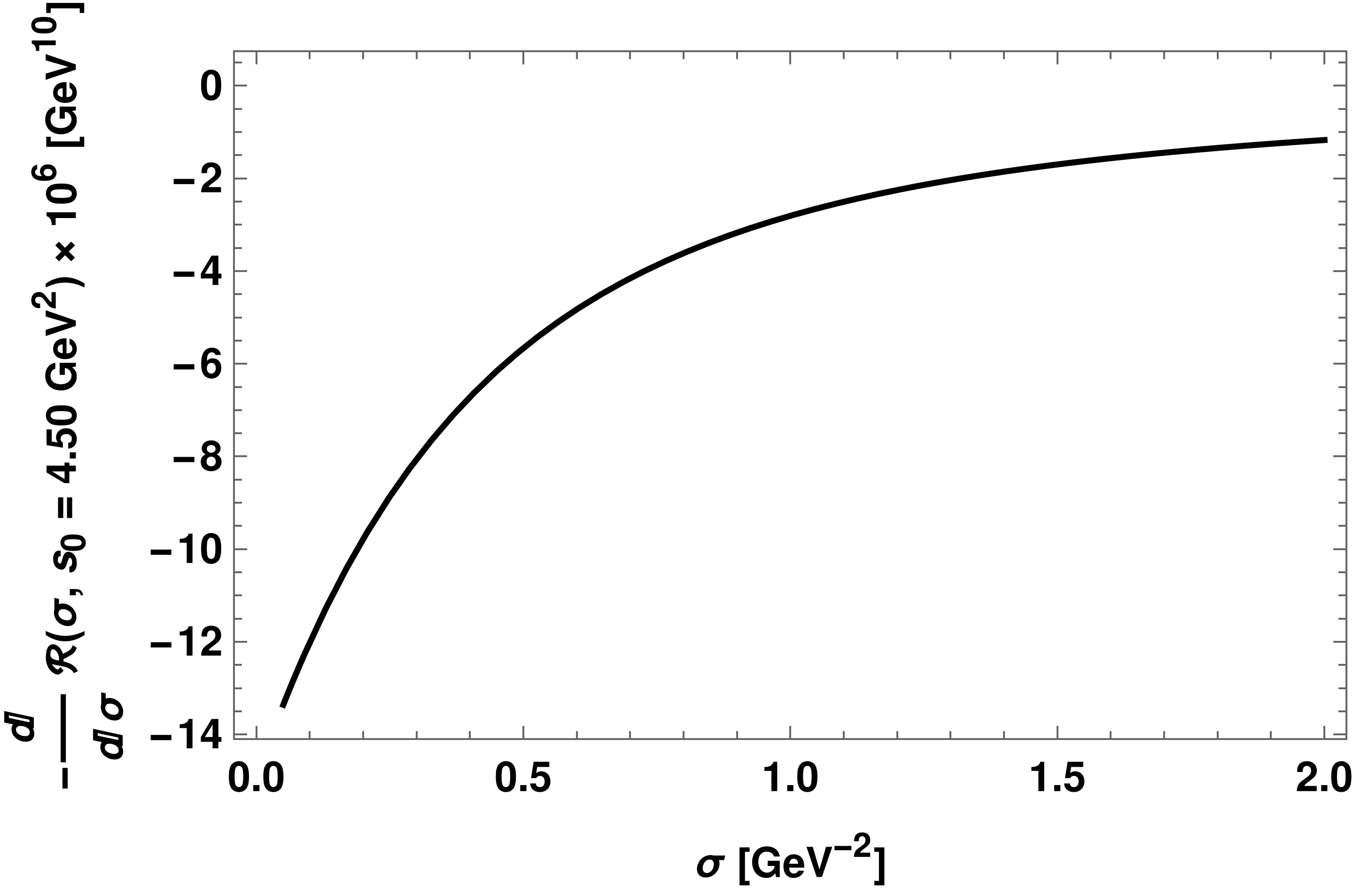}
    \caption{The order-1 (subtracted) LSR.}
    \label{faafo:b}
\end{subfigure}
\caption{The order-0 and order-1 (subtracted) LSRs of $J_3$ ($J_7$) from~\cite{Fu:2018ngx} 
at continuum threshold parameter $s_0 = 4.50~\text{GeV}^2$, the optimized value of 
$s_0$ determined in~\cite{Fu:2018ngx}.  
Both LSRs are negative at Borel parameter $\sigma = 0.265~\text{GeV}^{-2}$,
the optimized value of $\sigma$ determined in~\cite{Fu:2018ngx}. 
Note that, in~\cite{Fu:2018ngx}, Laplace sum rules are denoted $M$ and the Borel parameter 
is denoted $\tau$.} 
\label{faafo}
\end{figure}

Our focus in this paper is $0^{+-}$, $\flav$ tetraquarks, 
denoted as $\tq$ following the classification scheme of Ref.~\cite{Gershon:2022xnn}.
In~\cite{Fu:2018ngx}, correlation functions of eight interpolating currents were 
considered at leading order (LO) in perturbation theory.
In the chiral limit, correlation functions of these eight currents 
are pairwise degenerate, corresponding to four independent currents.
Of these four, only two were identified as leading to Laplace sum-rules (LSRs) 
with converging operator product expansion (OPE) series, and, of these two,
only one led to an LSRs analysis stable against deviations from vacuum saturation.
This current, $J_3$ (or, equivalently, $J_7$) in the notation of~\cite{Fu:2018ngx}, 
was analyzed using LSRs resulting in a mass prediction of 1.39~GeV
at optimized continuum threshold parameter $s_0 = 4.50~\text{GeV}^2$
and Borel parameter $\sigma = 0.265~\text{GeV}^{-2}$~\cite{Fu:2018ngx}.
Because sum rules for two-point functions relate an integrated 
QCD-predicted $\rho^{\text{(OPE)}}(t)$ 
to an integrated positive-valued hadronic spectral function $\rho(t)$
(see~(\ref{ulsr})--(\ref{ugsr}) below),
QCD sum rules must be positive to be physically consistent 
with an integrated hadronic spectral function 
(see e.g., Refs.~\cite{Becchi:1980vz,Narison:1981ts,Steele:1998ry,Lellouch:1997hp} 
that use the physical positivity constraint to obtain QCD sum-rule mass bounds on light quarks).
However, as can be seen in Fig.~\ref{faafo}, the LSRs of $J_3$ ($J_7$) from~\cite{Fu:2018ngx} 
are negative and, therefore, unphysical at the optimized $s_0$ and $\sigma$ values; 
hence, the corresponding mass prediction is not reliable.

Focusing on the current $J_7$ from~\cite{Fu:2018ngx} (see~\eqref{j1} below),
we extended the QCD sum-rules analysis of $\tq$ tetraquarks 
to include next-to-leading-order (NLO) QCD contributions to perturbation theory.
For several $0^{++}$, light-quark tetraquark currents, it has been shown that NLO contributions
to perturbation theory are surprisingly large~\cite{Groote:2014pva}.
The effects of NLO perturbation theory on a QCD sum-rules analysis of a
$0^{++}$, light-quark tetraquark current are explored in~\cite{Cid-Mora:2022kgu}.
It is therefore interesting to study whether light-quark, exotic-$J^{PC}$ tetraquarks
have similarly large NLO effects.
Furthermore, as the NLO perturbative contributions are necessarily positive, 
they could potentially fix the negative, unphysical LSRs of Fig.~\ref{faafo}.

The NLO diagrams 
that contribute to the \tq\ diagonal correlator defined in~(\ref{correlator})--(\ref{Pi_zero}) 
below are shown in Fig.~\ref{diagrams_nlo}.
Each diagram has four loops and contains nonlocal divergences.
Integrals were regulated using dimensional regularization, and nonlocal divergences 
were eliminated through diagrammatic 
renormalization~\cite{Hepp:1966eg,Zimmermann:1969jj,bogoliubov,collins} 
using the methodology for sum rules developed in Ref.~\cite{deOliveira:2022eeq}.
As discussed in~\cite{deOliveira:2022eeq}, diagrammatic renormalization is 
particularly convenient for radiative corrections to tetraquark correlation functions 
as it circumvents the problem of composite-operator mixing under renormalization. 
Also, it provides helpful consistency checks as nonlocal divergences are eliminated
diagram-by-diagram.

Rather than evaluating integrals analytically, we evaluated them using pySecDec, 
a program that numerically calculates dimensionally regularized 
integrals~\cite{Borowka:2017idc}.
pySecDec makes use of FORM~\cite{Vermaseren:2000nd,Kuipers:2013pba,Ruijl:2017dtg},
GSL~\cite{galassi}, and the CUBA library~\cite{Hahn:2004fe,Hahn:2014fua}.
It has been demonstrated that pySecDec can be successfully incorporated 
into the QCD sum-rules methodology at LO using a $0^{-+}$ charmonium hybrid 
current as an example~\cite{Esau:2018gdp}. 
In this paper, we demonstrate that pySecDec numerical loop-integration methods combined with diagrammatic renormalization techniques can be successfully implemented at NLO, establishing new calculational methods for higher-loop corrections in QCD sum-rules.

For the \tq\ diagonal correlator~(\ref{correlator})--(\ref{Pi_zero}) below,
we found that NLO contributions are large relative to LO perturbation theory.
This is similar to what was found for $0^{++}$, light-quark
tetraquarks~\cite{Groote:2014pva,Cid-Mora:2022kgu}.
To assess the importance of the NLO corrections to physical predictions, 
we computed NLO perturbative contributions to Laplace, Gaussian, and finite-energy sum rules.
Using Gaussian sum rules (GSRs), we motivated a lower bound on the $\tq$ tetraquark ground state mass, $M$,
and using LSRs, we determined an upper bound on $M$. 
Omitting NLO perturbation theory, we found that $2.4~\text{GeV}\lesssim M \leq 4.6~\text{GeV}$
contrary to the predictions of~\cite{Fu:2018ngx}.
The discrepancy is due to our analysis being restricted to positive, physical sum rules.
Including both LO and NLO perturbation theory, we found that $2.2~\text{GeV}\lesssim M \leq 4.2~\text{GeV}$.  
With NLO perturbation theory, the resulting mass scale was lowered somewhat, 
but we still found no evidence for a \tq\ state lighter than 1.9~GeV, even when taking into
account uncertainties in QCD parameters.
Furthermore, it is worth noting that
this mass range is comparable to the QCD sum-rule mass prediction for $0^{+-}$, light-quark
hybrids~\cite{Ho:2018cat}, suggesting the possibility of hybrid-tetraquark mixing.

\section{Next-to-Leading-Order Perturbation Theory}
We investigate $\tq$ tetraquarks using the current
\begin{equation}
\label{j1}
    J_\mu = u_a^TCd_b\big(\bar{u}_a\gamma_\mu C\bar{d}_b^{\,T}
    - \bar{u}_b\gamma_\mu C\bar{d}_a^{\,T}\big)
    - u_a^T C\gamma_\mu d_b\big(\bar{u}_aC\bar{d}_b^{\,T}
    - \bar{u}_bC\bar{d}_a^{\,T}\big),
\end{equation}
denoted $J_7$ in~\cite{Fu:2018ngx},
with charge conjugation operator $C$, quark colour indices $a$ and $b$, and massless $u$ and $d$ quarks.
As discussed in Ref.~\cite{Fu:2018ngx}, this current couples to different isospin multiplets, 
but because our calculations do not include isospin-breaking effects, 
our conclusions concerning \tq\ masses are isospin-degenerate.
The diagonal correlator of~(\ref{j1}) is
\begin{equation}
\begin{aligned}
    \Pi_{\mu \nu}(q)
    & = i\int\!\mathrm{d}^4 x\,
    e^{iq\cdot x} \langle\Omega|T j_{\mu}(x)j^{\dagger}_{\nu}(0)|\Omega\rangle\\
    & = q_\mu q_\nu\Pi^{\text{(S)}}(q^2)
    + \left(q_\mu q_\nu - q^2 g_{\mu\nu}\right)\Pi^{\text{(V)}}(q^2)
\label{correlator}
\end{aligned}
\end{equation}
where $\Pi^{\text{(S)}}(q^2)$ and $\Pi^{\text{(V)}}(q^2)$ probe $0^{+-}$ and $1^{--}$ states respectively.
We focus on $\Pi^{\text{(S)}}(q^2)$ where
\begin{equation}\label{Pi_zero}
  \Pi^{\text{(S)}}(q^2) = \frac{q_{\mu}q_{\nu}}{q^4}\Pi_{\mu\nu}(q^2).    
\end{equation}
We omit the superscript ``(S)'' from $\Pi^{\text{(S)}}(q^2)$ from here on.  
As discussed above, at LO, the LSRs based on this current have good OPE convergence 
properties and stability under variations in QCD parameter inputs~\cite{Fu:2018ngx}.
For the QCD sum-rules analyses of Section~\ref{sumrules}, we actually only need the imaginary
part of $\Pi$; thus, for convenience, we define
\begin{equation*}\label{rho_defn}
  \rho(t) = \lim_{\delta\to 0^{+}}\frac{\Pi(t+\ii\delta)-\Pi(t-\ii\delta)}{2\pi\ii}=
  \frac{1}{\pi} \Im\Pi(t).
\end{equation*}

We calculate $\rho(t)$ within the OPE~\cite{Wilson:1969zs,collins} in which 
perturbation theory, $\rho^{\text{(pert)}}(t)$, is supplemented by nonperturbative condensate terms, 
$\rho^{\text{(cond)}}(t)$,
\begin{equation}\label{rho_pert_cond}
  \rho(t) \rightarrow \rho^{\text{(OPE)}}(t) 
   = \rho^{\text{(pert)}}(t) + \rho^{\text{(cond)}}(t).
\end{equation}
In the chiral limit of massless $u$ and $d$ quarks, we consider LO, $\rho^{\text{(LO)}}(t)$,  
and NLO, $\rho^{\text{(NLO)}}(t)$, contributions to $\rho^{\text{(pert)}}(t)$ \ie\
\begin{equation}\label{rho_lo_plus_nlo}
 \rho^{\text{(pert)}}(t) = \rho^{\text{(LO)}}(t) + \rho^{\text{(NLO)}}(t)
\end{equation}
 where~\cite{Fu:2018ngx}
\begin{equation}\label{rho_lo}
    \rho^{(\text{LO})}(t) = \frac{t^3}{61440\pi^6}.
\end{equation}
Taking into account condensates up to and including a mass dimension of six
(\ie\ 6d), we have~\cite{Fu:2018ngx}
\begin{equation}\label{rho_cond}
  \rho^{\text{(cond)}}(t) = \frac{t}{1536\pi^5} \langle\alpha G^2\rangle 
    - \frac{\kappa}{12\pi^2} \langle\bar{q}q\rangle^2
\end{equation}
where the 4d gluon condensate value used is~\cite{Narison:2011rn}
\begin{equation}
  \langle\alpha G^2 \rangle = (0.075 \pm 0.02)\ \text{GeV}^4
\end{equation}
and the 3d quark condensate value used is~\cite{Shifman:1978bx,Launer:1983ib} 
\begin{equation}
  \langle\bar{u}u\rangle =  \langle\bar{d}d\rangle \equiv
  \langle\bar{q}q\rangle = -(0.23 \pm 0.03 )^3~\text{GeV}^3.
\end{equation}
As discussed in~\cite{Fu:2018ngx}, the chiral limit of 
massless quarks and $SU(2)$ flavour-symmetric QCD condensate
corrections imply that the $\tq$ predictions emerging from $\Pi(q^2)$ will be isopsin-degenerate.
The parameter $\kappa$ in~\eqref{rho_cond} quantifies deviations from the vacuum saturation hypothesis, 
with $\kappa=1$ corresponding to vacuum saturation~\cite{Shifman:1978bx,Shifman:1978by}.
However, there is considerable evidence that vacuum saturation underestimates the 
6d condensates, and so, consistent with Refs.~\cite{Narison:2002woh,Bertlmann:1987ty,Narison:1995jr},
we use $\kappa=2$ as our central value with $\kappa=3$ as an upper bound. 
(See Ref.~\cite{Gubler:2018ctz} for a recent review of QCD condensate determinations.)

\begin{figure}[hbt]
\centering
  \begin{subfigure}{0.3\textwidth}
  \centering
  \includegraphics[scale=0.5]{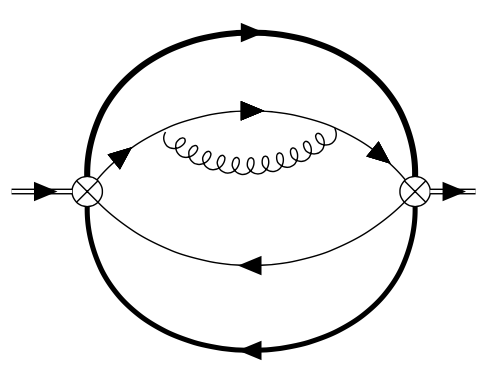}
  \caption{Self-energy diagram (SE).\newline}
  \label{diagrams_nlo_a}
  \end{subfigure}
  \hfill
  \begin{subfigure}{0.3\textwidth}
  \centering
   \includegraphics[scale=0.5]{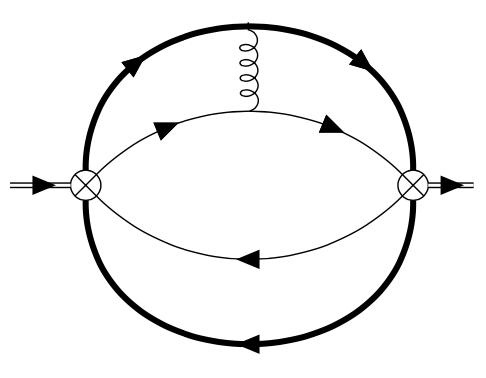}

  \caption{Type 1 gluon-exchange diagram (EX1).}
  \label{diagrams_nlo_b}
  \end{subfigure}
  \hfill
  \begin{subfigure}{0.3\textwidth}
  \centering
    \includegraphics[scale=0.5]{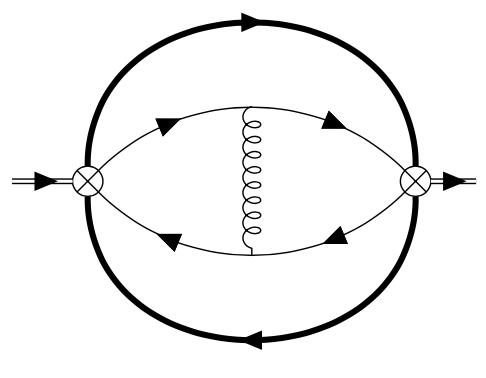}

  \caption{Type 2 gluon-exchange diagram (EX2).}
  \label{diagrams_nlo_c}
  \end{subfigure}
  \hfill
\caption{The NLO perturbative diagrams of $\Pi\left(q^2\right)$.  
  The $\otimes$ denotes the Feynman rule for the current~\eqref{j1}.
  Thin lines are $u$ quarks. Thick lines are $d$ quarks.}
\label{diagrams_nlo}
\end{figure}

The diagrams that contribute to $\Pi(q^2)$ at NLO are shown in Fig.~\ref{diagrams_nlo}, and $\rho^{\text{(NLO)}}(t)$ is then  extracted from these diagrams.
The self-energy diagram (SE) shown in Fig.~\ref{diagrams_nlo_a} has a multiplicity of four
as the gluon line can be attached to any of the (massless) quark lines.
In the gluon-exchange diagram shown in Fig.~\ref{diagrams_nlo_b}, 
the gluon line connects two quark lines oriented in the same direction.
We call this an exchange diagram of Type~1 (EX1) and note that it has a multiplicity of two.
In the gluon-exchange diagram shown in Fig.~\ref{diagrams_nlo_c}, 
the gluon line connects two quark lines oriented in opposite directions.
We call this an exchange diagram of Type~2 (EX2) and note that it has a multiplicity of four.

 All diagrams of Fig.~\ref{diagrams_nlo} contain nonlocal divergences that must be eliminated.
 Regularization is handled using dimensional regularization in $D=4+2\epsilon$ dimensions
 at minimal subtraction (MS) renormalization scale $\mu$.
 We renormalize each diagram using diagrammatic renormalization as discussed in Ref.~\cite{deOliveira:2022eeq}.
At NLO, diagrammatic renormalization (see \eg\
Refs.~\cite{Hepp:1966eg,Zimmermann:1969jj,bogoliubov,collins}) 
first requires isolation of the subdivergences arising from the
one-loop subdiagram(s) of an individual bare NLO diagram. 
Counterterm diagrams generated from the subdivergences are then calculated and 
added to the bare diagram to obtain the renormalized diagram. 
The process is repeated for all bare diagrams, and the final result is the 
renormalized correlation function with the coupling identified as $\alpha_s(\nu)$ 
at renormalization scale $\nu$ in the chosen scheme. 
Advantages of the diagrammatic approach include an increase in computational efficiency, 
particularly when conventional renormalization would result in a large operator-mixing basis 
(such as tetraquark systems with a basis of approximately 10 operators~\cite{Narison:1983kn}). 
Ref.~\cite{deOliveira:2022eeq} also shows how the diagrammatic method can be conceptually 
understood in terms of conventional operator renormalization.  
In summary, the diagrammatic renormalization process~\cite{deOliveira:2022eeq} requires that,
for each diagram, subdiagrams that lead to nonlocal divergences are identified, and
their MS divergences are isolated.
 For each subdivergence isolated, a counterterm vertex having the opposite value is defined.
 New counterterm diagrams that include counterterm vertices are added to the
 original diagram yielding a result free of nonlocal divergences.  
 A novel aspect of this paper is the implementation of diagrammatic renormalization 
 methodology via numerical loop integration methods using pySecDec~\cite{Borowka:2017idc} 
 as outlined below.
 
The diagram of Fig.~\ref{diagrams_nlo_a} contains a subdivergence from the quark self-energy.
The counterterm corresponding to the one-loop quark self-energy is well-known
(see~\cite{pascualandtarrach} for example).
For massless quarks, 
\begin{equation}\label{SE_counterterm}
\begin{tikzpicture} [baseline=(b)]
\begin{feynman} [horizontal=a to b]
  \vertex [label=above:\(a{,}\,i\)] (a);
  \vertex [right=0.75cm of a] (b);
  \vertex [right=2cm of b] (c);
  \vertex [right=0.75cm of c, label=above:\(b{,}\,j\)] (d);
  \diagram* {
    (a) -- [fermion, momentum'=\(p\)] (b) -- [fermion] (c) -- [fermion] (d),
    (b) -- [gluon, half left] (c)
  };
\end{feynman}
\end{tikzpicture}
\quad\Longrightarrow\quad
\begin{tikzpicture} [baseline=(b)]
\begin{feynman} [horizontal=a to b]
  \vertex [label=above:\(a{,}\,i\)] (a);
  \node [right=1.5cm of a, square dot, label=above right:\(1\)] (b);
  \vertex [right=1.5cm of b, label=above:\(b{,}\,j\)] (c);
  \diagram* {
    (a) -- [fermion, momentum'=\(p\)] (b) -- [fermion] (c),
  };
\end{feynman}
\end{tikzpicture}
= \frac{i g_s^2}{12\pi^2\epsilon} \delta^{ba} \slashed{p}_{ji}.
\end{equation}
We represent a counterterm vertex as a $\blacksquare$ labelled by an integer indicating which
subdivergence it corresponds to.
In~(\ref{SE_counterterm})--(\ref{EX2_counterterm}), 
the indices $\{a,\ldots,\,d\}$ represent quark colour whereas $\{i,\ldots,\,k\}$ are Dirac indices.
The diagram of Fig.~\ref{diagrams_nlo_b} contains two divergent subdiagrams
each comprising the gluon line, a current insertion, and the two quark lines that connect them.
We find
\begin{equation}
\begin{tikzpicture}[baseline=(a)]
\begin{feynman}[horizontal=a to b]
  \vertex (a);
  \node[right=1cm of a, crossed dot, label=above left:\(\mu\)] (b);
  \vertex[above=1.75cm of b, label=above:\(a{,}\,i\)] (c);
  \vertex[below=1.75cm of b, label=below:\(d{,}\,\ell\)] (d);
  \vertex[above right=1.41cm of b]  (e);
  \vertex[above right=2.47cm of b, label=above right:\(b{,}\,j\)] (f);
  \vertex[above=1cm of b] (g);
  \vertex[below right=2.47cm of b, label=below right:\(c{,}\,k\)] (h);
  \diagram* {
    (a) -- [double distance=0.3mm, with arrow=6mm] (b),
    (d) -- [fermion, line width=0.6mm] (b) -- [fermion, line width=0.6mm] (g) -- [fermion, line width=0.6mm] (c),
    (h) -- [fermion] (b) -- [fermion] (e) -- [fermion] (f),
    (e) -- [gluon] (g)
  };
\end{feynman}
\end{tikzpicture}
\Longrightarrow\quad
\begin{tikzpicture}[baseline=(a)]
\begin{feynman}[horizontal=a to b]
  \vertex (a);
  \node[right=1cm of a, square dot, label=above left:\(\mu\), label=right:\(2\)] (b);
  \vertex[above=1.75cm of b, label=above:\(a{,}\,i\)] (c);
  \vertex[below=1.75cm of b, label=below:\(d{,}\,\ell\)] (d);
  \vertex[above right=2.47cm of b, label=above right:\(b{,}\,j\)] (f);
  \vertex[below right=2.47cm of b, label=below right:\(c{,}\,k\)] (h);
  \diagram* {
    (a) -- [double distance=0.3mm, with arrow=6mm] (b),
    (d) -- [fermion, line width=0.6mm] (b) -- [fermion, line width=0.6mm] (c),
    (h) -- [fermion] (b) -- [fermion] (f)
  };
\end{feynman}
\end{tikzpicture}
= \Gamma^{\text{(EX1)}}_{\mu}
\end{equation}
where
\begin{equation}\label{EX1_counterterm}
  \Gamma^{\text{(EX1)}}_{\mu} =
  \frac{g_s^2}{24\pi^2\epsilon} \left(\delta^{ad}\delta^{bc} - \delta^{ac}\delta^{bd}\right) 
  \Big(4 C_{ij}\big(C\gamma_\mu\big)_{\ell k}- \big(\gamma_\mu C\big)_{ij}C_{\ell k}\Big).
\end{equation}
The diagram of Fig.~\ref{diagrams_nlo_c} contains two divergent subdiagrams, again,
each comprising the gluon line, a current insertion, and the two quark lines that connect them.
We find
\begin{equation}
\begin{tikzpicture}[baseline=(a.base)]
\begin{feynman}[horizontal=a to b]
  \vertex (a);
  \node[right=1cm of a, crossed dot, label=above left:\(\mu\)] (b);
  \vertex[above=1.75cm of b, label=above:\(a{,}\,i\)] (c);
  \vertex[below=1.75cm of b, label=below:\(d{,}\,\ell\)] (d);
  \vertex[above right=1.41cm of b]  (e);
  \vertex[above right=2.47cm of b, label=above right:\(b{,}\,j\)] (f);
  \vertex[below right=1.41cm of b] (g);
  \vertex[below right=2.47cm of b, label=below right:\(c{,}\,k\)] (h);
  \diagram* {
    (a) -- [double distance=0.3mm, with arrow=6mm] (b),
    (d) -- [fermion, line width=0.6mm] (b) -- [fermion, line width=0.6mm] (c),
    (h) -- [fermion] (g) -- [fermion] (b) -- [fermion] (e) -- [fermion] (f),
    (e) -- [gluon] (g)
  };
\end{feynman}
\end{tikzpicture}
\Longrightarrow\quad
\begin{tikzpicture}[baseline=(a.base)]
\begin{feynman}[horizontal=a to b]
  \vertex (a);
  \node[right=1cm of a, square dot, label=above left:\(\mu\), label=right:\(3\)] (b);
  \vertex[above=1.75cm of b, label=above:\(a{,}\,i\)] (c);
  \vertex[below=1.75cm of b, label=below:\(d{,}\,\ell\)] (d);
  \vertex[above right=2.47cm of b, label=above right:\(b{,}\,j\)] (f);
  \vertex[below right=2.47cm of b, label=below right:\(c{,}\,k\)] (h);
  \diagram* {
    (a) -- [double distance=0.3mm, with arrow=6mm] (b),
    (d) -- [fermion, line width=0.6mm] (b) -- [fermion, line width=0.6mm] (c),
    (h) -- [fermion] (b) -- [fermion] (f)
  };
\end{feynman}
\end{tikzpicture}
= \Gamma^{\text{(EX2)}}_{\mu}
\end{equation}
where
\begin{equation}\label{EX2_counterterm}
\Gamma^{\text{(EX2)}}_{\mu} = 
  \frac{g_s^2}{384\pi^2\epsilon}
  \left(5\delta^{ad}\delta^{bc} + \delta^{ac}\delta^{bd}\right)
  \Big(\big(\gamma^\rho\gamma^\sigma C\big)_{ij} \big(C\gamma_\mu\gamma_\rho\gamma_\sigma\big)_{\ell k} 
- \big(\gamma_\mu \gamma^\rho\gamma^\sigma C\big)_{ij}\big(C\gamma_\rho\gamma_\sigma)_{\ell k}\big)\Big).
\end{equation}

\begin{figure}[hbt]
\centering
  \begin{subfigure}{0.3\textwidth}
  \centering
  \includegraphics[scale=0.5]{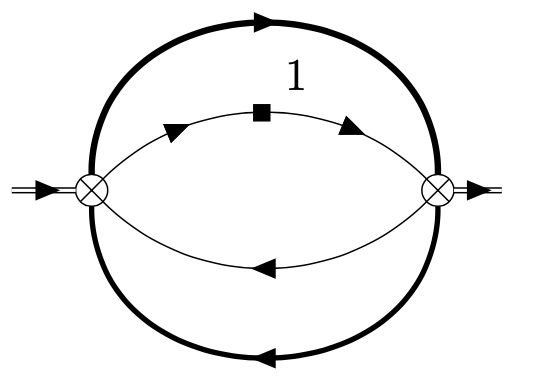}
  \caption{Self-energy counterterm diagram (SEC).}
  \label{counterterm_diagrams_nlo_a}
  \end{subfigure}
  \hfill
  \begin{subfigure}{0.3\textwidth}
  \centering
  \includegraphics[scale=0.5]{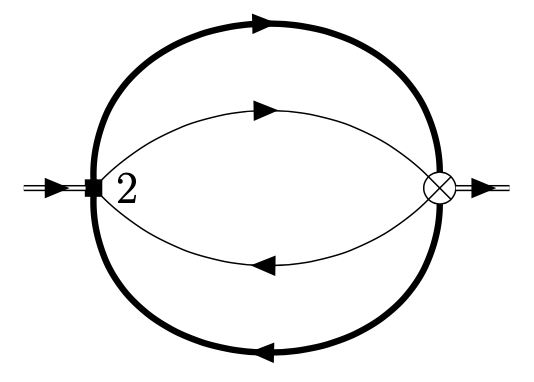}
  \caption{Type 1 gluon-exchange counterterm diagram (EXC1).}
  \label{counterterm_diagrams_nlo_b}
  \end{subfigure}
  \hfill
  \begin{subfigure}{0.3\textwidth}
  \centering
    \includegraphics[scale=0.5]{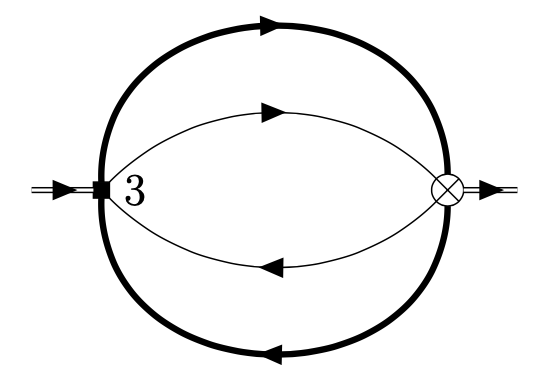}
  \caption{Type 2 gluon-exchange counterterm diagram (EXC2)}
  \label{counterterm_diagrams_nlo_c}
  \end{subfigure}
  \hfill
\caption{The counterterm diagrams needed to eliminate nonlocal divergences from 
  the diagrams of Fig.~\ref{diagrams_nlo}.
  The $\otimes$ denotes the Feynman rule for the current~\eqref{j1}.
  The $\blacksquare$ denotes a counterterm vertex.
  Thin lines are $u$ quarks. Thick lines are $d$ quarks.}
\label{counterterm_diagrams_nlo}
\end{figure}

The counterterm diagrams needed to eliminate nonlocal divergences 
from the diagrams of Fig.~\ref{diagrams_nlo}
are shown in Fig.~\ref{counterterm_diagrams_nlo}.
The self-energy counterterm diagram (SEC) of Fig.~\ref{counterterm_diagrams_nlo_a} has a multiplicity of four.
The Type~1 gluon-exchange counterterm diagram (EXC1) 
of Fig.~\ref{counterterm_diagrams_nlo_b} has a multiplicity of four,
the multiplicity of the diagram of Fig.~\ref{diagrams_nlo_b} multiplied by two as the
counterterm vertex can replace either current insertion.
The Type~2 gluon-exchange counterterm diagram (EXC2)
of Fig.~\ref{counterterm_diagrams_nlo_c} has a multiplicity of eight,
the multiplicity of the diagram of Fig.~\ref{diagrams_nlo_c} multiplied by two.

\begin{figure}[ht]
\centering
\begin{tabular}{cc}
    \includegraphics[scale=0.56]{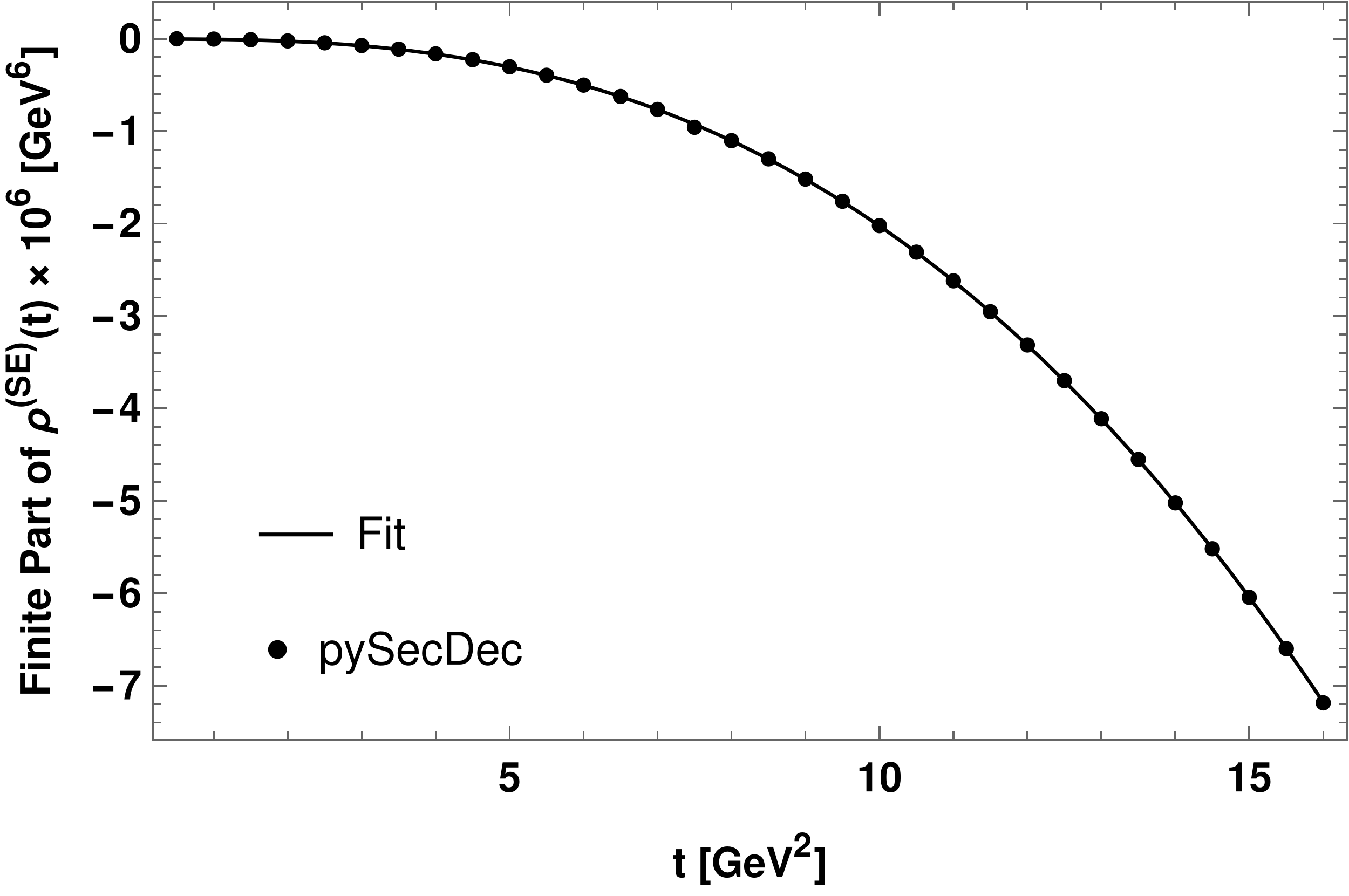} & \includegraphics[scale=0.56]{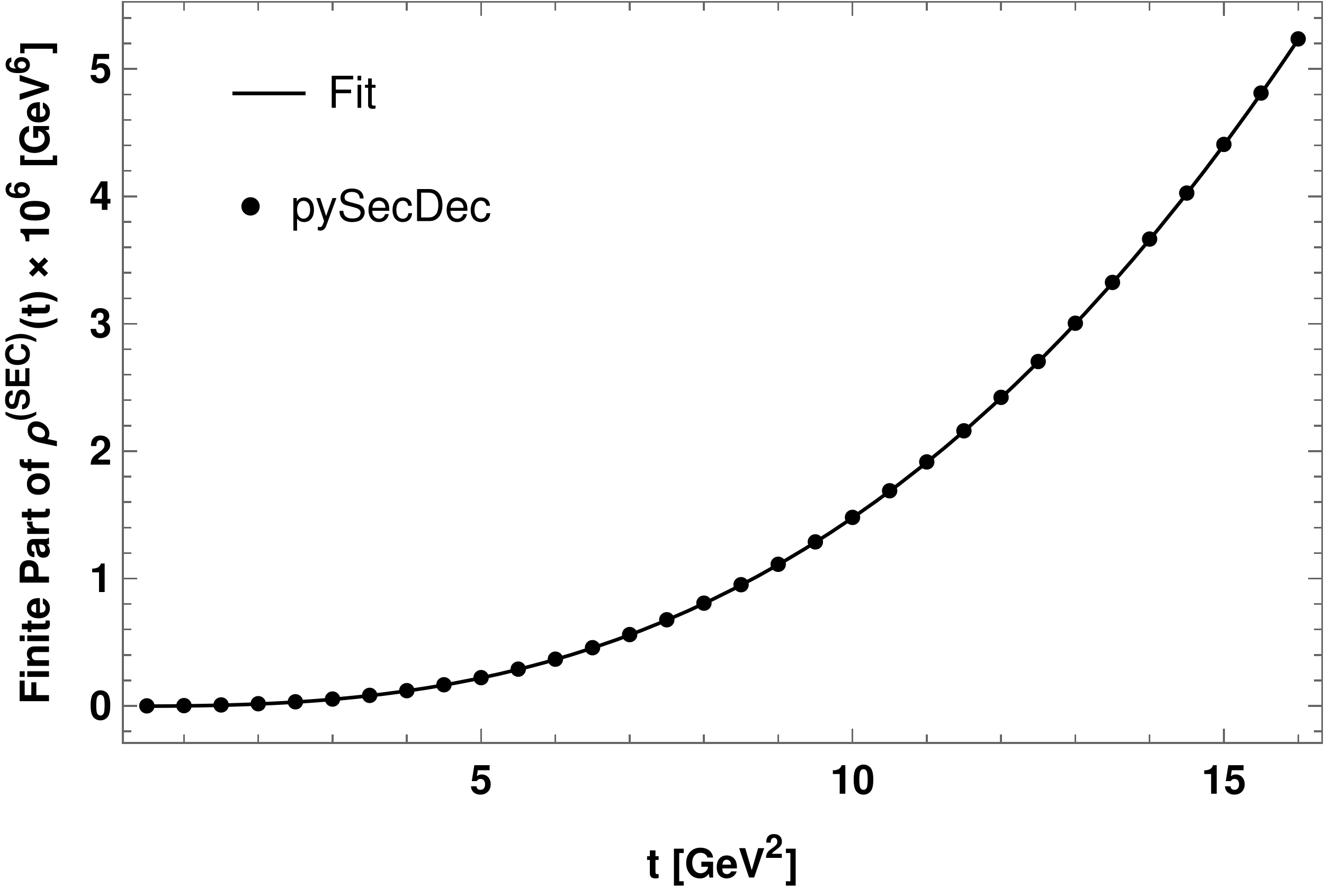} \\
    \includegraphics[scale=0.56]{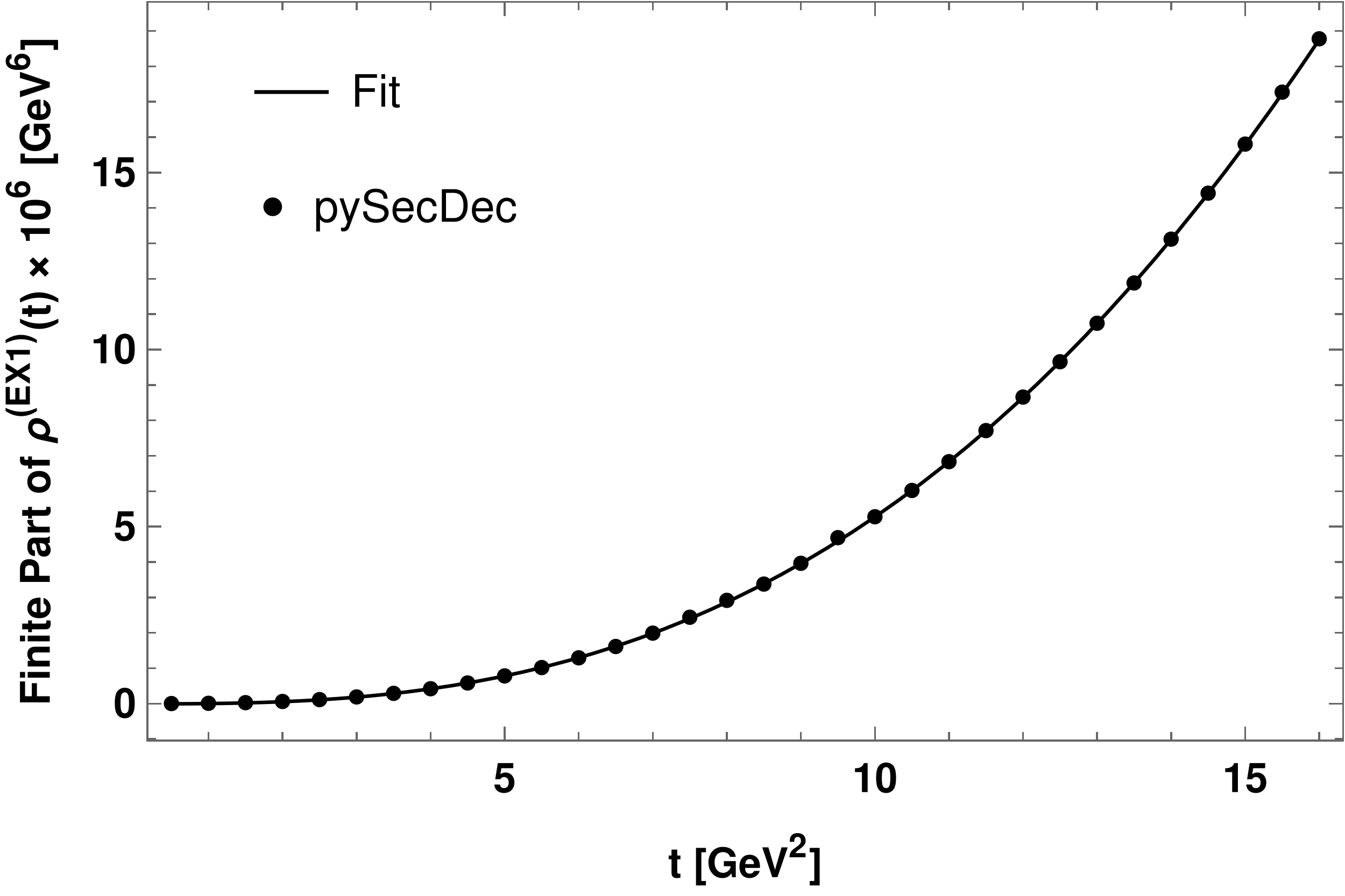} & \includegraphics[scale=0.56]{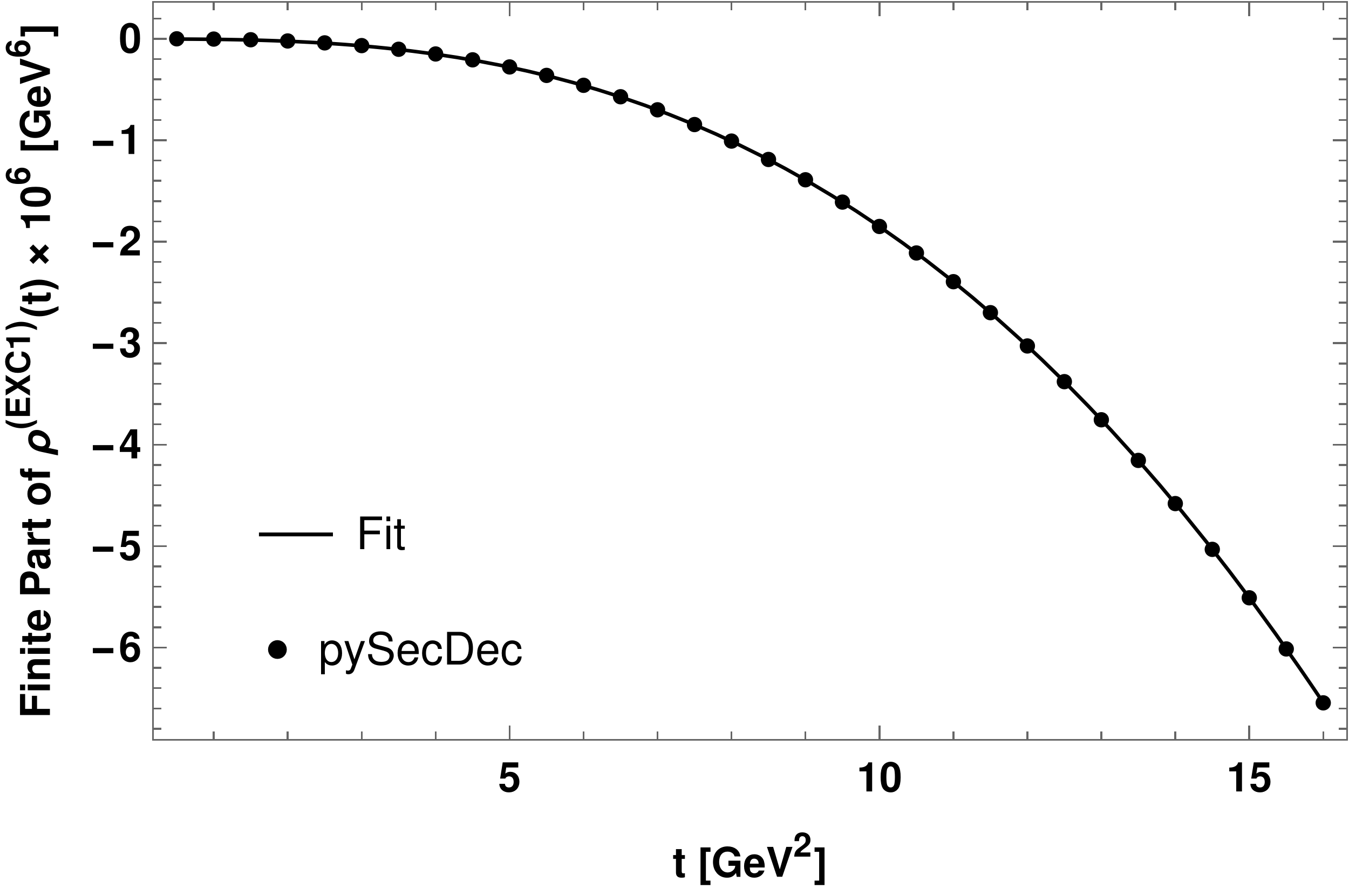} \\
    \includegraphics[scale=0.56]{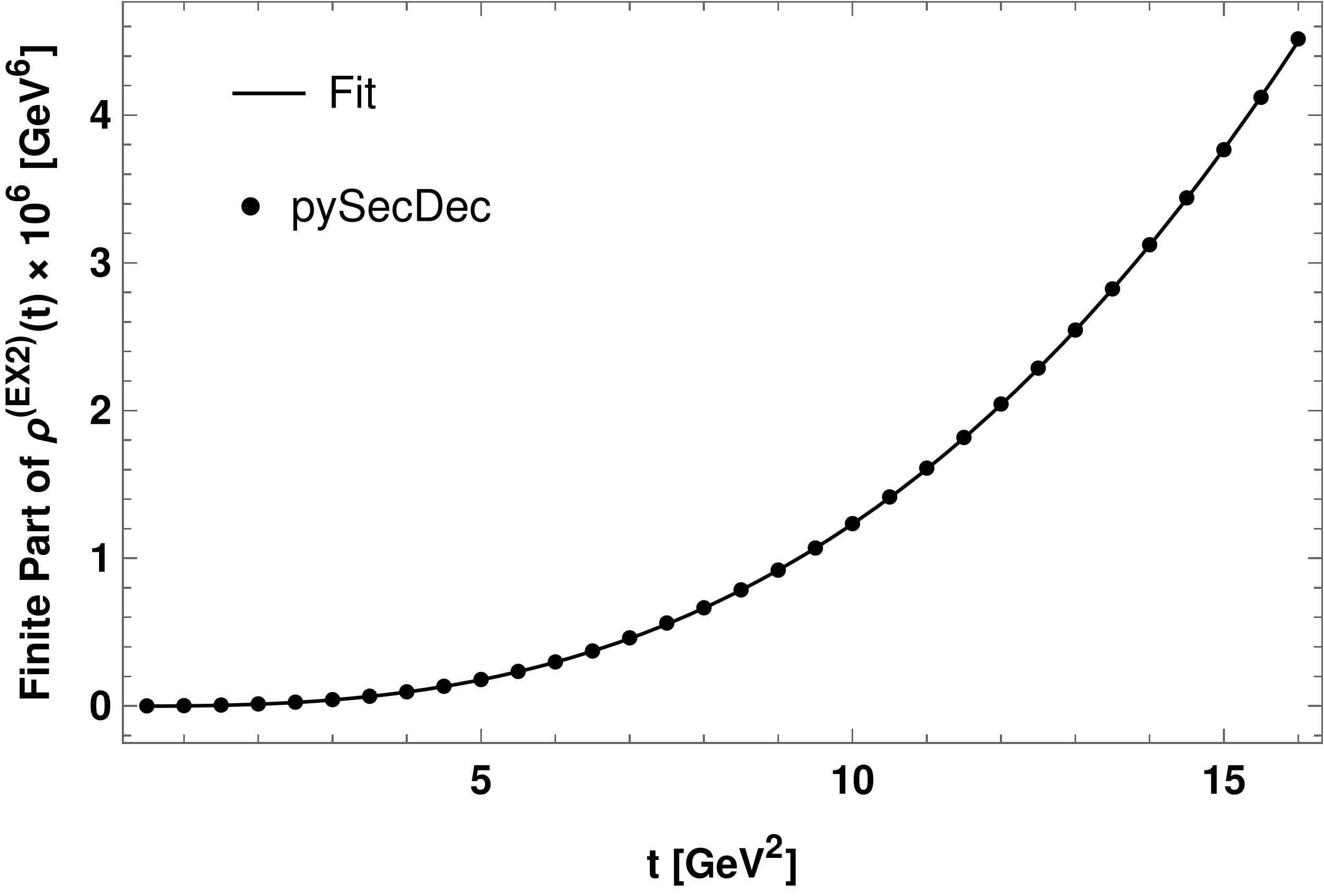} & \includegraphics[scale=0.56]{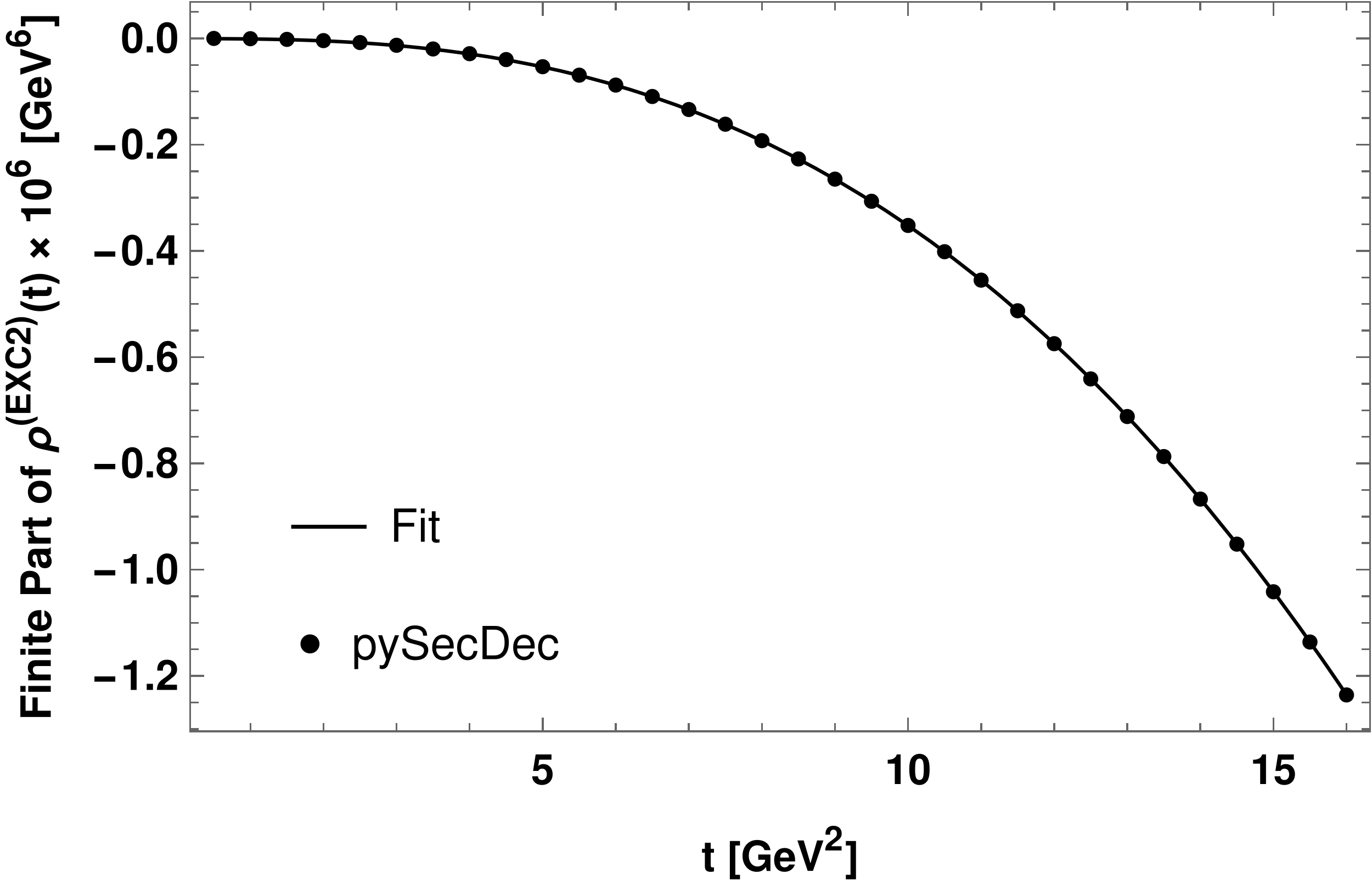} \\
\end{tabular}
\caption{Fits (solid lines) of the finite parts of~(\ref{rho_form}) to pySecDec-generated data (dots). 
  The error bars due to pySecDec numerical uncertainties are much smaller than the dots.}
\label{rho_fits}
\end{figure}

We denote a particular NLO diagram from Fig.~\ref{diagrams_nlo} or Fig.~\ref{counterterm_diagrams_nlo}
with a superscript (A) where
$A\in\{\text{SE},\ \text{EX1},\ \text{EX2},\ \text{SEC},\ \text{EXC1},\ \text{EXC2}\}$.
Then, including multiplicities,
\begin{equation}\label{nlo_breakdown}
    \rho^{(\text{NLO})}(t) = 4\rho^{\text{(SE)}}(t)
    +4\rho^{\text{(SEC)}}(t)+2\rho^{\text{(EX1)}}(t)
    +4\rho^{\text{(EXC1)}}(t)+4\rho^{\text{(EX2)}}(t)+8\rho^{\text{(EXC2)}}(t).
\end{equation}
As the $u$ and $d$ quarks are massless, each $\rho^{\text{(A)}}(t)$ takes the form
\begin{equation}\label{rho_form}
    \rho^{\text{(A)}}(t) 
    = g_s^2 t^3 \left(\frac{a^{\text{(A)}}}{\epsilon} + b^{\text{(A)}} + c^{\text{(A)}} L\right)
\end{equation}
where $a^{\text{(A)}}$, $b^{\text{(A)}}$, and $c^{\text{(A)}}$ are constants and where
\begin{equation}
    L = \log\!\left(\frac{t}{\mu^2}\right).
\end{equation}
Using pySecDec, we numerically evaluated the imaginary parts of all six NLO diagrams 
(excluding the $g_s^2$ factors) over a range of values of $t$ at $\mu  = 1~\text{GeV}$. 
The values of $a^{\text{(A)}}$ were easily identified as the coefficients of 
$\epsilon^{-1}$ in the resulting data.
We extracted values of $b^{\text{(A)}}$ and $c^{\text{(A)}}$ 
by fitting the finite parts of~(\ref{rho_form}), \ie\ the terms free of $\epsilon$, 
to the coefficients of $\epsilon^{0}$ in the data.
As a benchmark of our methodology, we first successfully reproduced~(\ref{rho_lo}) using this pySecDec method  
before calculating NLO corrections.
In Fig.~\ref{rho_fits}, for each NLO diagram, we plot the fitted finite part of~(\ref{rho_form})
along with the coefficient of $\epsilon^0$ in the pySecDec-generated data.
In all cases, there is excellent agreement between the data and the fitted function, 
and the theoretical uncertainty in the coefficients arising from the fitting procedure is negligible.

\begin{figure}[ht]
\centering
\begin{tabular}{cc}
    \includegraphics[scale=0.58]{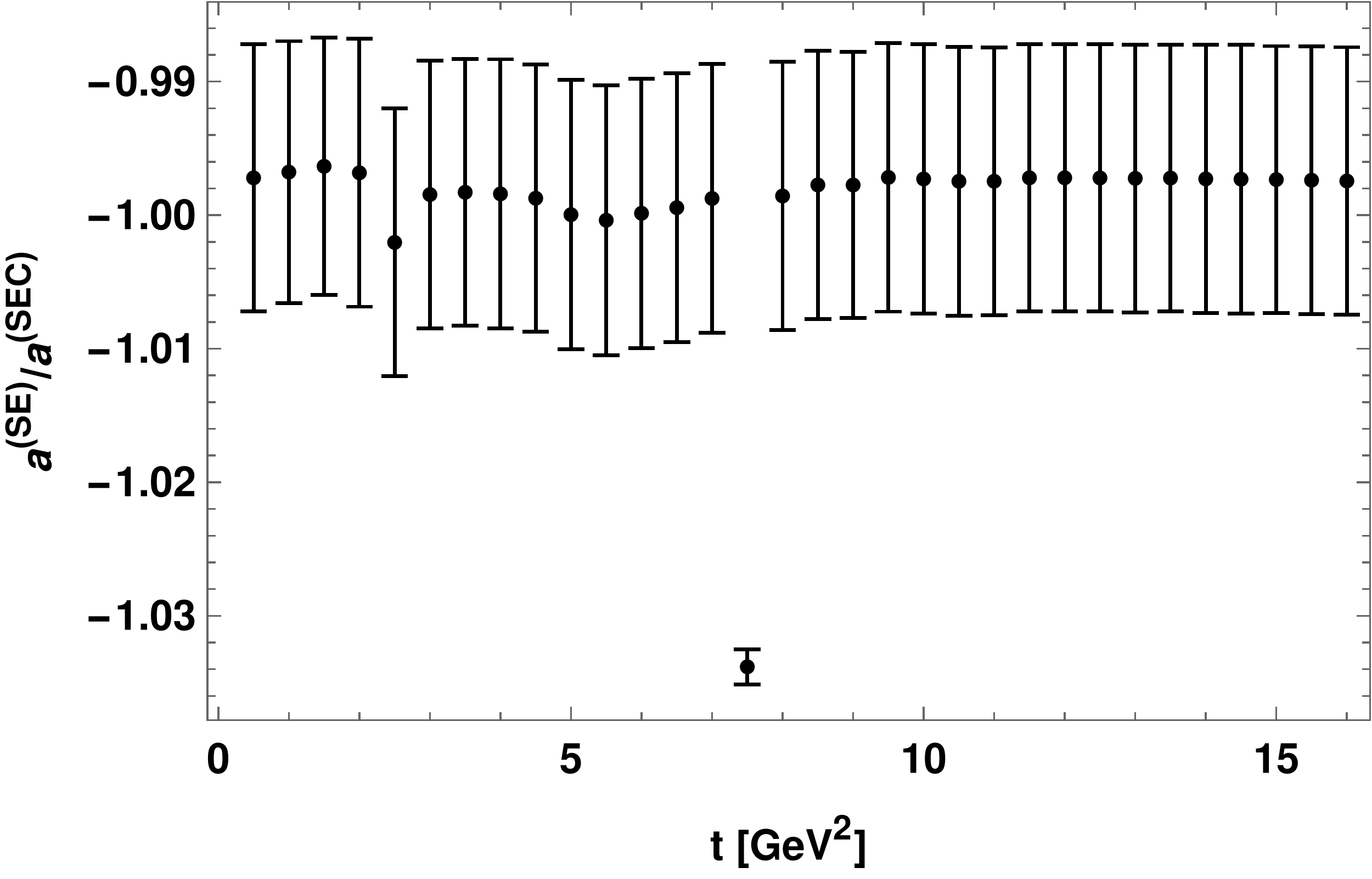} & \includegraphics[scale=0.58]{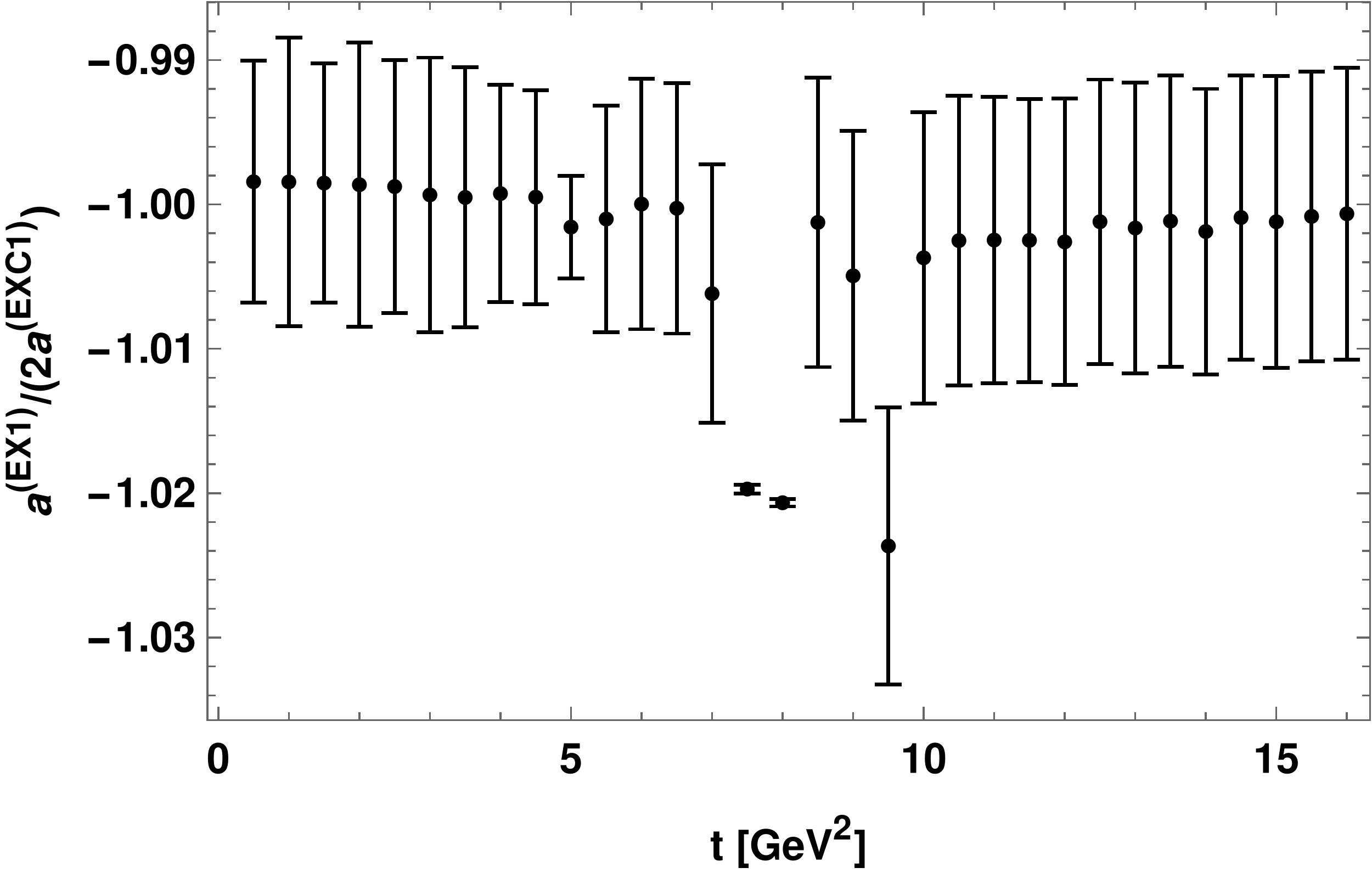} \\
     \includegraphics[scale=0.58]{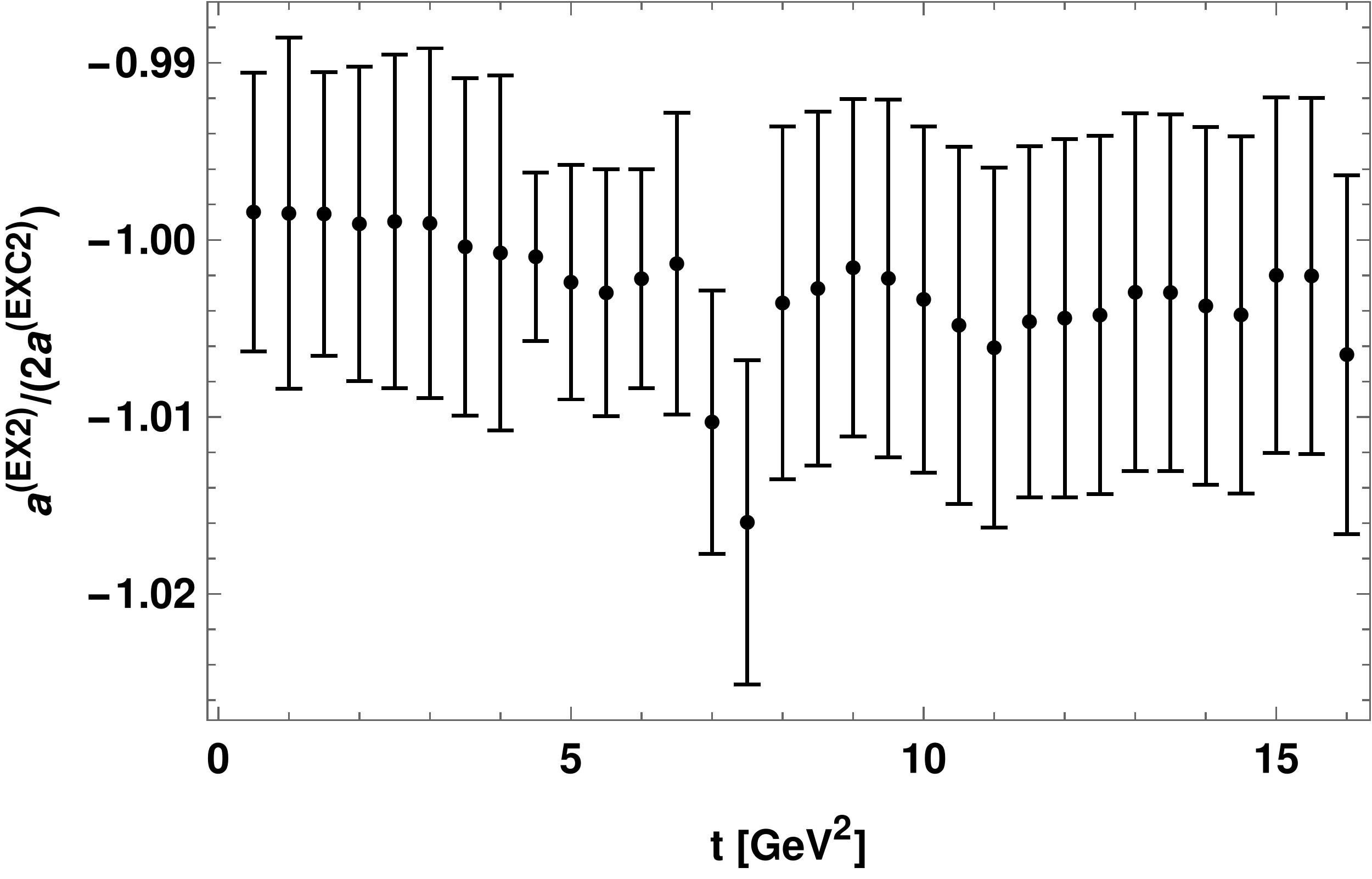} & 
\end{tabular}
\caption{Ratios of the divergent parts of diagrams and their corresponding counterterm diagrams. 
  Error bars correspond to numerical uncertainties estimated by pySecDec.}
\label{rho_ratios}
\end{figure}

The sum of a diagram and its counterterm diagrams must be free of nonlocal divergences
implying, here, that the various divergent parts, \ie\ the $\epsilon^{-1}$ terms,
of~(\ref{rho_form}) must cancel in pairs. Therefore,
\begin{gather}
    a^{\text{(SE)}} = - a^{\text{(SEC)}}\label{se_ratio}\\
    a^{\text{(EX1)}} = - 2a^{\text{(EXC1)}}\label{ex1_ratio}\\
    a^{\text{(EX2)}} = - 2a^{\text{(EXC2)}}.\label{ex2_ratio}
\end{gather}
The factors of two in~(\ref{ex1_ratio}) and~(\ref{ex2_ratio}) are due to the two possible
locations of the counterterm vertex in Fig.~\ref{counterterm_diagrams_nlo_b} 
and Fig.~\ref{counterterm_diagrams_nlo_c} respectively.
In Fig.~\ref{rho_ratios}, we plot $a^{\text{(SE)}}/a^{\text{(SEC)}}$, 
$a^{\text{(EX1)}}/(2a^{\text{(EXC1)}})$, 
and $a^{\text{(EX2)}}/(2a^{\text{(EXC2)}})$ using values of $a^{\text{(A)}}$ obtained from fitting.
Within numerical uncertainty, each ratio is consistent with a constant value of -1
in excellent agreement with~(\ref{se_ratio})--(\ref{ex2_ratio}).
There are, however, a handful of outliers that violate~(\ref{se_ratio})--(\ref{ex2_ratio}) by 
a few percent.
As the $u$ and $d$ quarks are massless, there is no special physical significance of any 
value of $t$, and so the outliers seem to be minor numerical anomalies. 
We speculate that the origin of these numerical anomalies is associated with our choice $\mu=1\,\gev$
corresponding to a modified minimal subtraction (\msbar) scale 
(see~\eqref{mu_ms_bar} below) of $\bar\mu^2=7.05\,\gev^2$ in close proximity to the values of $t$ at 
the anomalies. 
It seems plausible that pySecDec could encounter numerical challenges at this scale because of 
the natural combination $1/\epsilon+\log(t/\mu^2)-\log(4\pi)+\gamma_E$ occurring in dimensional regularization.
From Fig.~\ref{rho_fits}, it is clear that the finite parts do not contain any such 
numerical anomalies.

For our final expression for $\rho^{\text{(NLO)}}(t)$, 
we transform from an MS to an \msbar\ result in order to make use of 
\msbar\ QCD quantities provided in Ref.~\cite{ParticleDataGroup:2020ssz}, for example.
With
\begin{equation}
\label{mu_ms_bar}
  \mu^2 = \frac{\ee^{\gamma_E}}{4\pi}\bar{\mu}^2,
\end{equation}
we have
\begin{equation}
  L = \bar{L} -\gamma_E + \log(4\pi)
\end{equation}
where
\begin{equation}
  \bar{L} = \log\!\left(\frac{q^2}{\bar{\mu}^2}\right)
\end{equation}
in terms of \msbar\ renormalization scale $\bar{\mu}$.
Then, ignoring the divergent parts of~(\ref{rho_form}) as the sum of all such contributions
has been shown to cancel in pairs, we find
\begin{align}
  \rho^{(\text{SE})}(t)   &= g_s^2 t^3 \left(-2.23\times10^{-9} + 5.76\times10^{-10} \bar{L}\right)\label{SE}\\
  \rho^{(\text{EX1})}(t)  &= g_s^2 t^3 \left(5.80\times10^{-9} - 1.49\times10^{-9} \bar{L}\right)\\
  \rho^{(\text{EX2})}(t)  &= g_s^2 t^3 \left(1.33\times10^{-9} - 2.87\times10^{-10} \bar{L}\right)\\
  \rho^{(\text{SEC})}(t)  &= g_s^2 t^3 \left(1.63\times10^{-9} - 4.29\times10^{-10} \bar{L}\right)\\
  \rho^{(\text{EXC1})}(t) &= g_s^2 t^3 \left(-2.04\times10^{-9} + 5.36\times10^{-10} \bar{L}\right)\\
  \rho^{(\text{EXC2})}(t) &= g_s^2 t^3 \left(-3.89\times10^{-10} + 1.07\times10^{-10} \bar{L}\right)\label{EXC2}.
\end{align}
Substituting~(\ref{SE})--(\ref{EXC2}) 
into~(\ref{nlo_breakdown})
gives
\begin{equation}\label{rho_nlo}
    \rho^{(\text{NLO})}(t) 
    = g_s^2 t^3 \left(3.30\times10^{-9}-5.40\times10^{-10}\bar{L}\right).
\end{equation}
Then, substituting~(\ref{rho_lo}) and~(\ref{rho_nlo}) 
into~(\ref{rho_lo_plus_nlo}) gives
\begin{equation}\label{rho_pert_final}
    \rho^{\text{(pert)}}(t) = (1.69\times 10^{-8})\,t^3 
    \left(1 + \alpha_s\left(2.45 - 0.401\bar{L}\right)\right)
\end{equation}
where
\begin{equation}
    \alpha_s=\frac{g_s^2}{4\pi}
\end{equation}
is the running strong coupling at the renormalization scale $\bar \mu$.
For four active flavours (\ie\ $n_f=4$) at one-loop order,
\begin{equation}\label{running_alpha}
    \alpha_s(\bar{\mu}) = \frac{\alpha_s(M_\tau)}
    {1 + \frac{25}{12\pi}\alpha_s(M_\tau)\log\left(\frac{\bar{\mu}^2}{M^2_{\tau}}\right)}
\end{equation}
where $M_{\tau}$, the $\tau$ mass, is 1.77~GeV and where 
$\alpha_s(M_{\tau}) = 0.330$~\cite{ParticleDataGroup:2020ssz}.

The perturbative results are consolidated by,
once again, expressing $\rho^{\text{(pert)}}(t)$ as 
(recall~(\ref{rho_lo_plus_nlo}))
\begin{equation}
\rho^{\text{(pert)}}(t) = \rho^{\text{(LO)}}(t) + \rho^{\text{(NLO)}}(t)
\label{tom_rho_expansion}
\end{equation}
where
\begin{equation}
 \rho^{\text{(LO)}}(t)= d_1 t^3\,,~
  \rho^{\text{(NLO)}}(t)= d_1 t^3\frac{\alpha_s}{\pi}  \left(d_2 + d_3\bar{L}\right)
  \label{tom_rho}
\end{equation}
implies
\begin{equation}
\label{rho_NLO}
    \rho^{\text{(pert)}}\triple{t}{\alpha_s}{\bar{\mu}} = 
    d_1 t^3  \bigg(1 + \frac{\alpha_s}{\pi} \big(d_2 + d_3\bar{L}\big)\bigg)
\end{equation}
with, from~(\ref{rho_pert_final}), 
\begin{equation}
d_1=1.69\times 10^{-8}\,, ~d_2=7.70\,, ~
d_3=-1.26.
\label{tom_dn}
\end{equation}
The NLO perturbative terms in~\eqref{rho_NLO} imply that $\rho(t)$ satisfies a
renormalization-group (RG) equation that contains an anomalous-dimension 
$\gamma_\rho(\alpha_s)$ contribution
\begin{equation}\label{gamma_rho}
 \Bigg( \bar\mu \frac{\partial}{\partial\bar\mu}+
 \beta\!\left(\alpha_s\right) \alpha_s \frac{\partial}{\partial\alpha_s}-2\gamma_\rho\!\left(\alpha_s\right)
 \Bigg)\,\rho(t) = 0
 \end{equation}
 where
 \begin{equation}
   \beta\left(\alpha_s\right)=\beta_1\frac{\alpha_s}{\pi}
   + \mathcal{O}\bigg(\frac{\alpha_s}{\pi}\bigg)^{\!2} 
   \text{ with } \beta_1=-\frac{11}{2}+\frac{n_f}{3}
 \end{equation}
 and
 \begin{equation}
   \gamma_\rho\left(\alpha_s\right) = \gamma_1\frac{\alpha_s}{\pi}
   + \mathcal{O}\bigg(\frac{\alpha_s}{\pi}\bigg)^{\!2}
   \text{ with }\gamma_1 = -d_3.
\end{equation}
However, up to NLO, the quantity 
\begin{equation}
\tilde\rho(t) = \alpha_s^{2\gamma_1/\beta_1} \rho(t)
\label{tilde_rho}
\end{equation}
satisfies an RG equation that does not contain an anomalous-dimension contribution, 
enabling standard RG approaches to QCD sum rules as discussed below.

\begin{figure}[htb]
\centering
\includegraphics[scale=0.8]{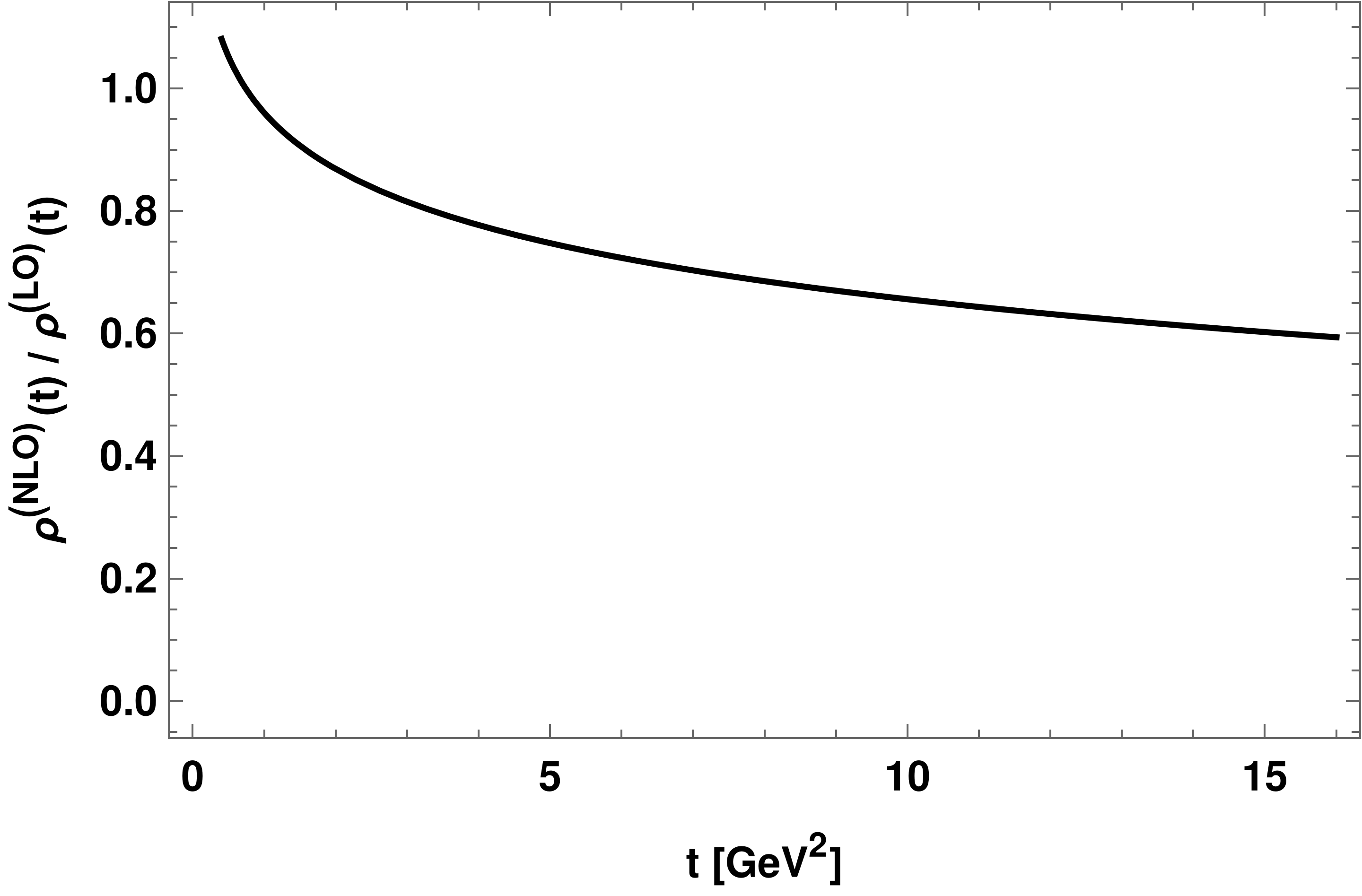}
\caption{The ratio of $\rho^{\text{(NLO)}}(t)$ to $\rho^{\text{(LO)}}(t)$
 (see Eq.~\eqref{tom_ratio})
for $\bar{\mu}=M_\tau$.
\label{rho_ratio}}
\end{figure}

\begin{figure}[htb]
\centering
\includegraphics[scale=0.8]{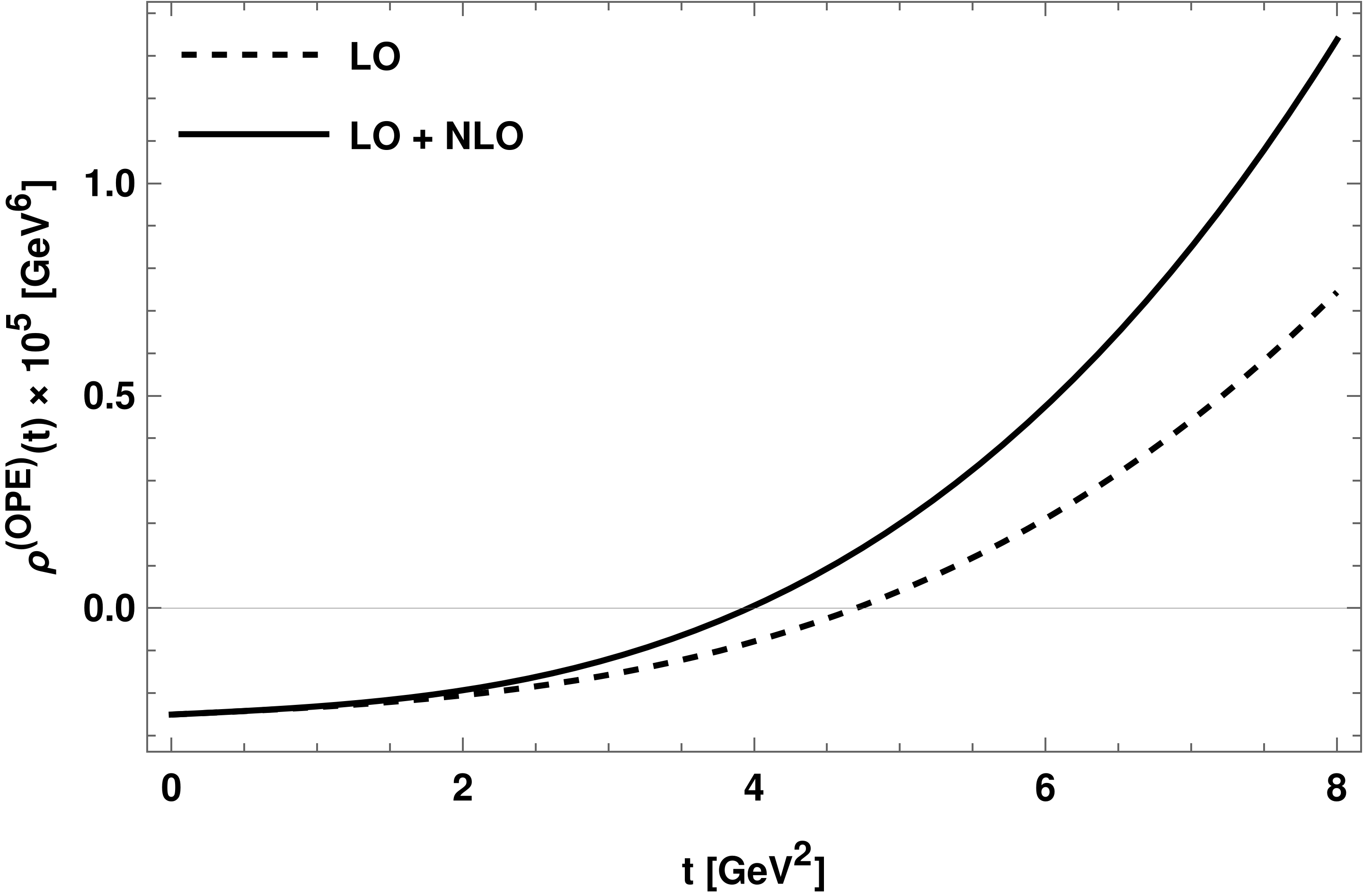}
\caption{$\rho^{\text{(OPE)}}(t)$ with NLO perturbation theory (the solid line) 
  and without (the dashed line) at $\bar{\mu}=M_{\tau}$.}
\label{rho_ope}
\end{figure}

The relative size of the LO and NLO terms in \eqref{tom_rho} can be examined through 
the ratio 
\begin{equation}
 \frac{\rho^{\text{(NLO)}}(t)}{\rho^{\text{(LO)}}(t)}=\frac{\alpha_s}{\pi} \left(d_2 + d_3\bar{L}\right).
 \label{tom_ratio}
\end{equation}
Hence, with $d_2\sim 10$ from \eqref{tom_dn}, it is expected that NLO effects 
could be important.
Recalling from~(\ref{rho_lo_plus_nlo}) and \eqref{tom_rho_expansion} that 
$\rho^{\text{(pert)}}(t) = \rho^{\text{(LO)}}(t) + \rho^{\text{(NLO)}}(t)$,
we plot
in Fig.~\ref{rho_ratio}  the ratio \eqref{tom_ratio}
 for 
a characteristic renormalization scale
$\bar{\mu}=M_{\tau}$.
Over the range of values of $t$ considered in the figure,
$\rho^{\text{(NLO)}}(t)$ is, on average, roughly  75\%
the size of $\rho^{\text{(LO)}}(t)$.
In Fig.~\ref{rho_ope}, we plot $\rho^{\text{(OPE)}}(t)$ 
(see~(\ref{rho_pert_cond}), (\ref{rho_cond}), and~(\ref{rho_pert_final}))
with and without $\rho^{\text{(NLO)}}(t)$ at $\bar{\mu}=M_{\tau}$.
For $t\lesssim 4~\text{GeV}^2$, $\rho^{\text{(OPE)}}(t) < 0$
due to the large magnitude, negative contribution from the 6d quark condensate term. 
However, the NLO contributions do mitigate the effect of the 6d condensates 
by extending the $\rho^{\text{(OPE)}}>0$ region to lower values of $t$ compared to LO.
(Note that the zeroes of $\rho^{\text{(OPE)}}(t)$ and $\tilde{\rho}^{\text{(OPE)}}(t)$ are the same.)
This behaviour of $\rho^{\text{(OPE)}}$ is similar to that seen for $1^{-+}$, light-quark tetraquarks
in~\cite{Chen:2008qw,Chen:2008ne}
and disfavours the existence of \tq\ states lighter than $\approx$2~GeV.
But, of course, hadronic predictions cannot be extracted directly from
the QCD-calculated $\rho^{\text{(OPE)}}(t)$, and so,
for a more rigorous analysis, we relate $\rho^{\text{(OPE)}}(t)$ to the 
channel's hadronic spectral function through QCD sum rules.

\section{Tetraquark Ground State Mass Bounds from QCD Sum Rules}
\label{sumrules}
At $Q^2=-q^2>0$, the correlator $\Pi$ defined in~(\ref{correlator})--(\ref{Pi_zero}) 
satisfies the dispersion relation
\begin{equation}\label{dispersion}
  \Pi(Q^2) = -Q^6\int_{t_0}^\infty \frac{\rho(t)}{t^3 (t + Q^2)}\dt + \ldots
\end{equation}
where $\rho(t)$ is the hadronic spectral function,
$t_0\approx 0$ is a threshold parameter
corresponding to the squared energy needed to create real constituents,
and $\cdots$ represents a polynomial in $Q^2$ (subtraction constants).
When $\Pi(Q^2)$ is computed through the OPE, the dispersion relation~(\ref{dispersion}) 
connects QCD to hadronic physics, \ie\ quark-hadron duality.

\begin{figure}[htb]
\centering
\includegraphics[scale=0.8]{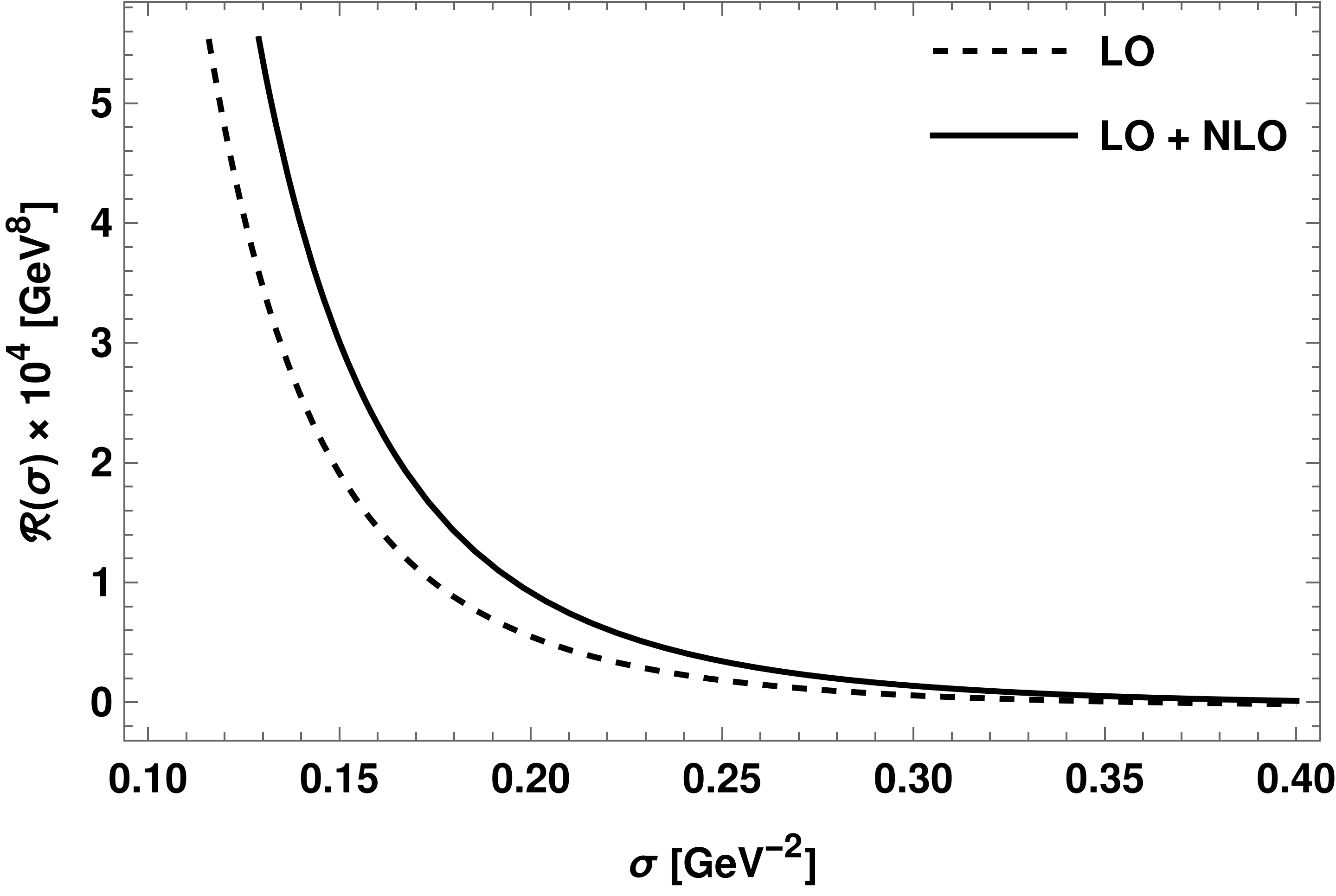}
\caption{The LSR $\lsr(\sigma)$ with NLO perturbation theory (the solid line) and 
  without (the dashed line).}
\label{lo_nlo_lsrs_plot}
\end{figure}

\begin{figure}[htb]
\centering
\includegraphics[scale=0.8]{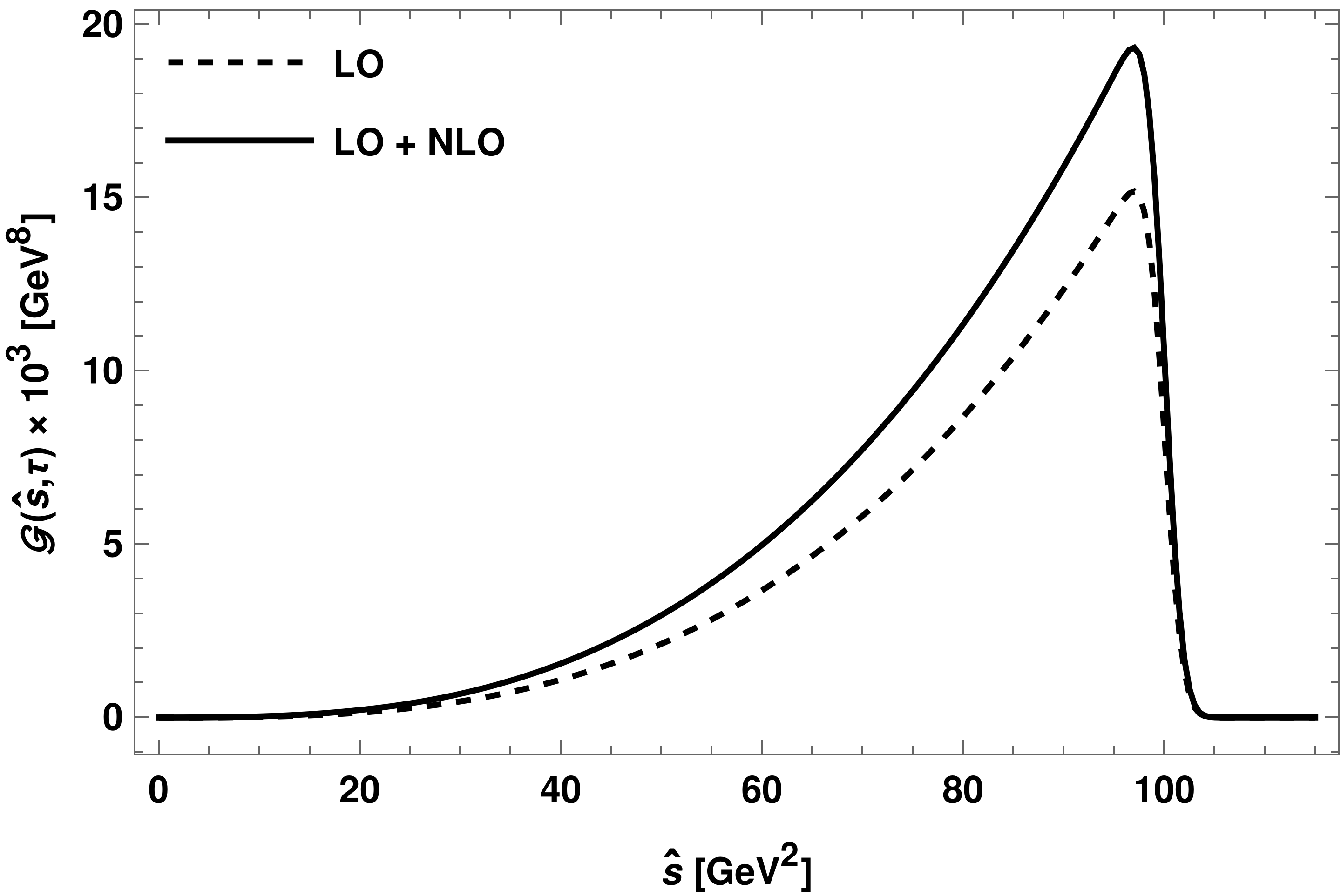}
\caption{The GSR $\gsr\double{\hat{s}}{\tau}$ with NLO perturbation theory (the solid line) and 
  without (the dashed line) at $\tau=1~\gev^4$.}
\label{lo_nlo_gsrs_plot}
\end{figure}

To reduce the contribution to the right-hand side (RHS) of~(\ref{dispersion})
from the high-energy behaviour of $\rho(t)$ (such as from excited states) 
as well as to eliminate subtraction constants and local field-theory divergences,
a transform is typically applied, leading to QCD sum rules.
Two examples of QCD sum rules are Laplace sum 
rules~\cite{Shifman:1978bx,Shifman:1978by,Reinders:1984sr,Narison:2002woh} and Gaussian sum 
rules~\cite{Bertlmann:1984ih,Orlandini:2000nv,Harnett:2000fy,Narison:2002woh}\footnote{Regarding 
arguments to sum rules, we follow the notation of~\cite{Bertlmann:1984ih}.}.
The (order-0) LSR, $\lsr(\sigma)$, is defined as
\begin{equation}\label{lsr_defn}
    \lsr(\sigma) = \frac{1}{\sigma}\,
    \lim_{N,\,Q^2\rightarrow\infty}\frac{(-Q^2)^N}{\Gamma(N)}
    \left(\frac{\mathrm{d}}{\mathrm{d}Q^2}\right)^N
    \!\Pi(Q^2)
\end{equation}
where $\sigma=\frac{N}{Q^2}$, the Borel parameter, is in GeV$^{-2}$.
The (order-0) GSR, $\gsr\double{\hat{s}}{\tau}$, 
is defined as
\begin{equation}\label{gsr_defn}
  \gsr\double{\hat{s}}{\tau} = \sqrt{\frac{\tau}{\pi}}\,
  \lim_{N,\,\Delta^2\rightarrow\infty}\frac{(-\Delta^2)^N}{\Gamma(N)}
  \left(\frac{\mathrm{d}}{\mathrm{d}\Delta^2}\right)^N
  \!\left(\frac{ \Pi(-\hat{s}-\ii\Delta) 
    -  \Pi(\hat{s}+\ii\Delta)}{\ii\Delta}\right)
\end{equation}
where $\tau=\frac{\Delta^2}{N}$ is in GeV$^4$.
Combining~(\ref{dispersion}) with $\Pi \rightarrow\Pi^{\text{(OPE)}}$ and~(\ref{lsr_defn}) 
gives (see~\cite{Shifman:1978bx,Bertlmann:1984ih} for details)
\begin{equation}\label{ulsr}
    \lsr(\sigma) 
    = \int_0^{\infty}\! \ee^{-\sigma t}\rho^{\text{(OPE)}}(t)\,\dt
    = \int_{0}^{\infty}\! \ee^{-\sigma t}\rho(t)\,\dt.
\end{equation}
Similarly, combining~(\ref{dispersion}) with $\Pi \rightarrow\Pi^{\text{(OPE)}}$
and~(\ref{gsr_defn}) gives (see~\cite{Bertlmann:1984ih,Orlandini:2000nv,Harnett:2000fy} for details)
\begin{equation}\label{ugsr}
    \gsr \double{\hat{s}}{\tau}
    = \frac{1}{\sqrt{4\pi\tau}}\int_{0}^{\infty}\! \ee^{-\frac{(t-\hat{s})^2}{4\tau}} \rho^{\text{(OPE)}}(t)\,\dt
    = \frac{1}{\sqrt{4\pi\tau}}\int_{0}^{\infty}\! \ee^{-\frac{(t-\hat{s})^2}{4\tau}} \rho(t)\,\dt.
\end{equation}
For massless quarks,
we set $\bar{\mu}=1/\sqrt{\sigma}$ in the LSR~\cite{Narison:1981ts} 
and $\bar{\mu}=\sqrt[4]{\tau}$ in the GSR~\cite{Bertlmann:1984ih,Orlandini:2000nv}. 
These RG-improvement results are based on an RG equation free of anomalous-dimension contributions;
hence, the anomalous-dimension factor $\alpha_s^{2\gamma_1/\beta_1}$ from~\eqref{tilde_rho} 
should be included as an additional positive multiplicative factor.  
However, the phenomenological analysis presented below is based on the sign of the GSR and an LSR ratio; 
in both cases, the anomalous-dimension factor has no effect and so can be ignored 
(see a similar argument in Ref~\cite{Cid-Mora:2022kgu}).
In Fig.~\ref{lo_nlo_lsrs_plot}, we plot $\lsr(\sigma)$, and,
in Fig.~\ref{lo_nlo_gsrs_plot}, we plot $\gsr\double{\hat{s}}{\tau}$ at 
$\tau=1~\gev^4$.
In both figures, the solid curve includes both LO and NLO perturbation theory 
whereas the dashed curve includes LO perturbation theory only.
It can be seen that NLO perturbation theory makes significant contributions to 
both the LSR and GSR.
Note that, for $\sigma\gtrsim 0.4~\text{GeV}^{-2}$, $\lsr(\sigma)$ becomes negative 
and, therefore, is unphysical for this region of Borel parameter.

We split the hadronic spectral function into hadronic and QCD continuum 
contributions
\begin{equation}\label{res_plus_cont}
  \rho(t) = \rho^{\text{(had)}}(t) + \rho^{\text{(OPE)}}(t)
  \theta(t-s_0)
\end{equation}
where $\rho^{\text{(had)}}(t)$ contains the resonance(s) content of $\rho(t)$,  
$s_0$ is the continuum threshold parameter, 
and $\theta(t-s_0)$ is a Heaviside step function.
Plugging~(\ref{res_plus_cont}) into~(\ref{ugsr}) gives
\begin{equation}\label{sgsr}
     \gsr \triple{\hat{s}}{\tau}{s_0}
    = \frac{1}{\sqrt{4\pi\tau}}\int_{0}^{s_0}\! \ee^{-\frac{(t-\hat{s})^2}{4\tau}} 
      \rho^{\text{(OPE)}}(t)\,\dt
    =  \frac{1}{\sqrt{4\pi\tau}}\int_{0}^{\infty}\! \ee^{-\frac{(t-\hat{s})^2}{4\tau}} 
      \rho^{\text{(had)}}(t)\,\dt
\end{equation}
where $\gsr\triple{\hat{s}}{\tau}{s_0}$ is a (continuum-)subtracted GSR.

\begin{figure}[htb]
\centering
\includegraphics[scale=0.8]{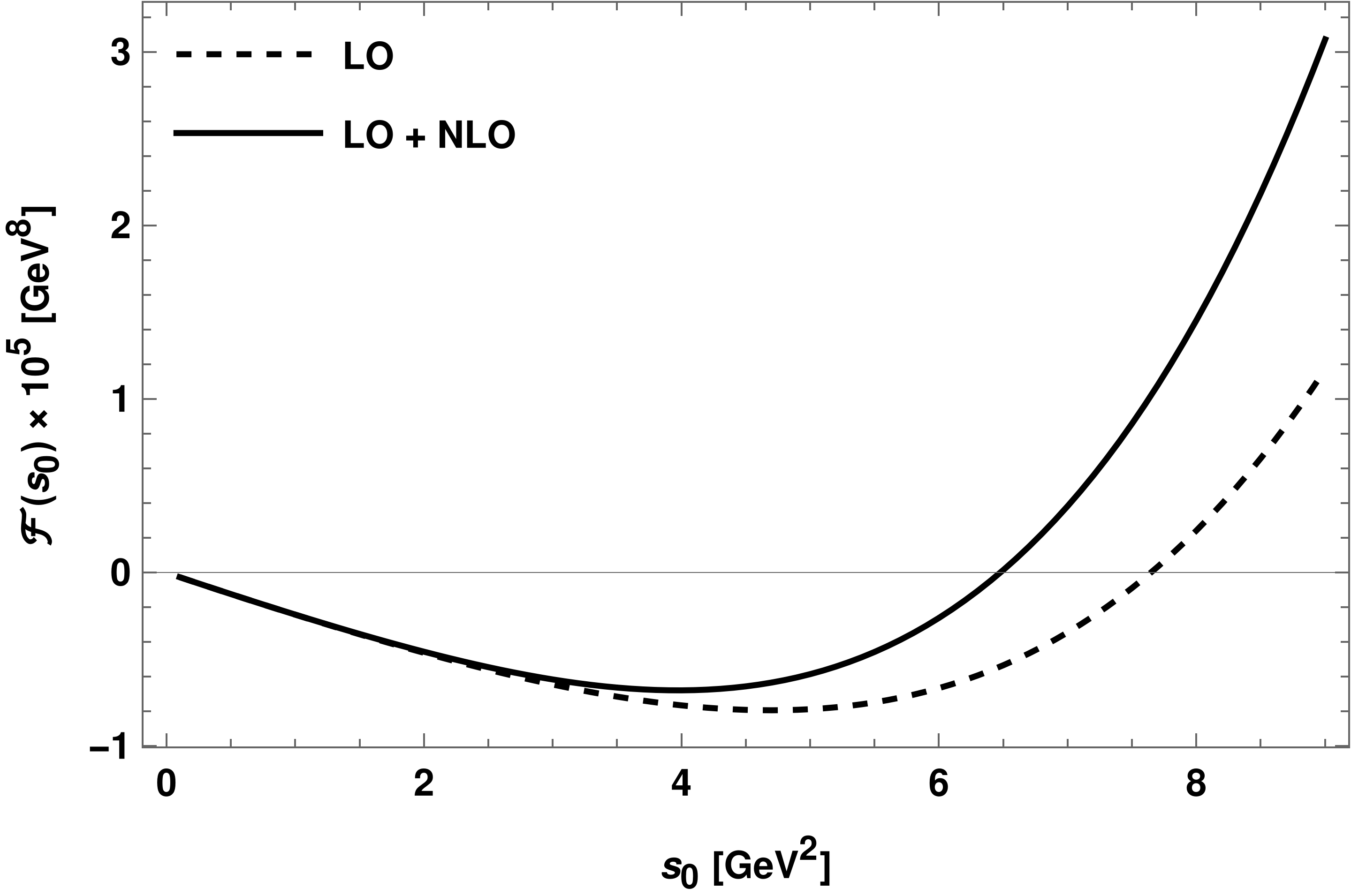}
\caption{The FESR $\mathcal{F}(s_0)$ with NLO perturbation theory (the solid line) 
  and without (the dashed line).}
\label{fesr_fig}
\end{figure}

Using~(\ref{sgsr}), we extract a lower bound on $s_0$. 
Integrating~(\ref{sgsr}) over $-\infty<\hat{s}<\infty$ gives
\begin{equation}\label{fesr}
     \mathcal{F} (s_0)
    = \int_{0}^{s_0}\! \rho^{\text{(OPE)}}(t)\,\dt
    =  \int_{0}^{\infty}\! \rho^{\text{(had)}}(t)\,\dt,
\end{equation}
a finite-energy sum rule (FESR).
In the FESR, we set $\bar{\mu}=\sqrt{s_0}$~\cite{Bertlmann:1984ih}.
As $\rho^{\text{(had)}}(t)\geq 0$, it follows that $\fesr(s_0)\geq 0$, 
and, as discussed above, omitting the anomalous-dimension factor has no effect on the sign of the FESR.
In Fig.~\ref{fesr_fig}, we plot $\fesr(s_0)$ with and without NLO perturbation theory.
Only values of $s_0$ that lead to positive $\fesr(s_0)$ are physically allowed.
Therefore, with NLO perturbation theory, we find that 
\begin{equation}\label{s0_bound_nlo}s_0 > 6.47~\text{GeV}^2,\end{equation} 
and, without NLO perturbation theory, we find that 
\begin{equation}\label{s0_bound_lo}s_0 > 7.64~\text{GeV}^2.\end{equation}
To reiterate, the bounds~(\ref{s0_bound_nlo}) and~(\ref{s0_bound_lo}) 
are constraints on $s_0$ that follow directly from the physical requirement that 
the FESR be positive. The optimized value $s_0 = 4.50~\text{GeV}^2$ used 
in~\cite{Fu:2018ngx} does not satisfy~(\ref{s0_bound_lo}).

\begin{figure}[htb]
\centering
\includegraphics[scale=0.8]{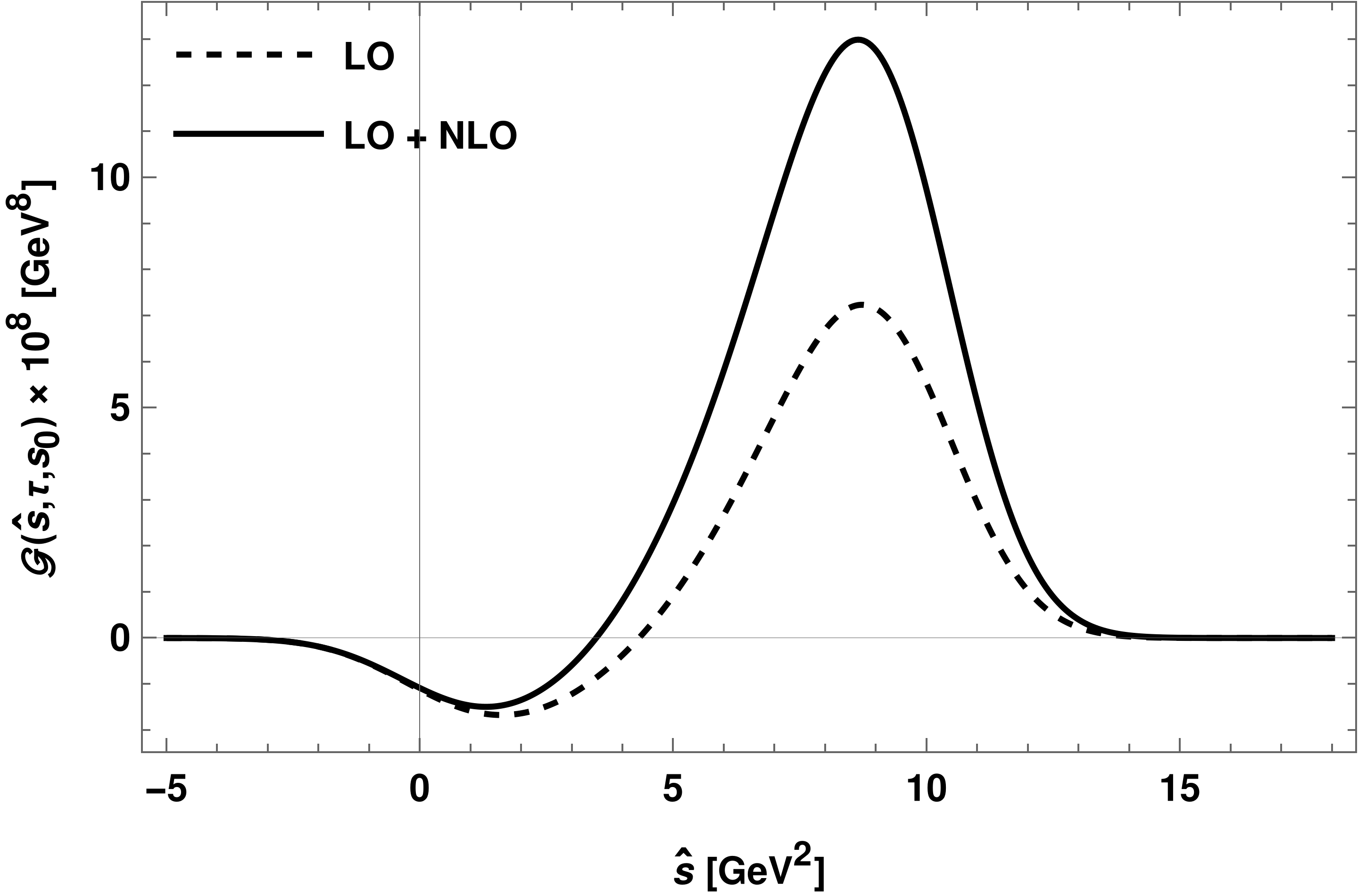}
\caption{The subtracted GSR $\gsr\triple{\hat{s}}{\tau}{s_0}$ at $\tau=1\,\text{GeV}^{4}$ 
  and $s_0=10\ \text{GeV}^2$ with NLO perturbation theory (the solid line) and without (the dashed line).}
\label{GSR_Plot}
\end{figure}

\begin{figure}[htb]
\centering
\includegraphics[scale=0.8]{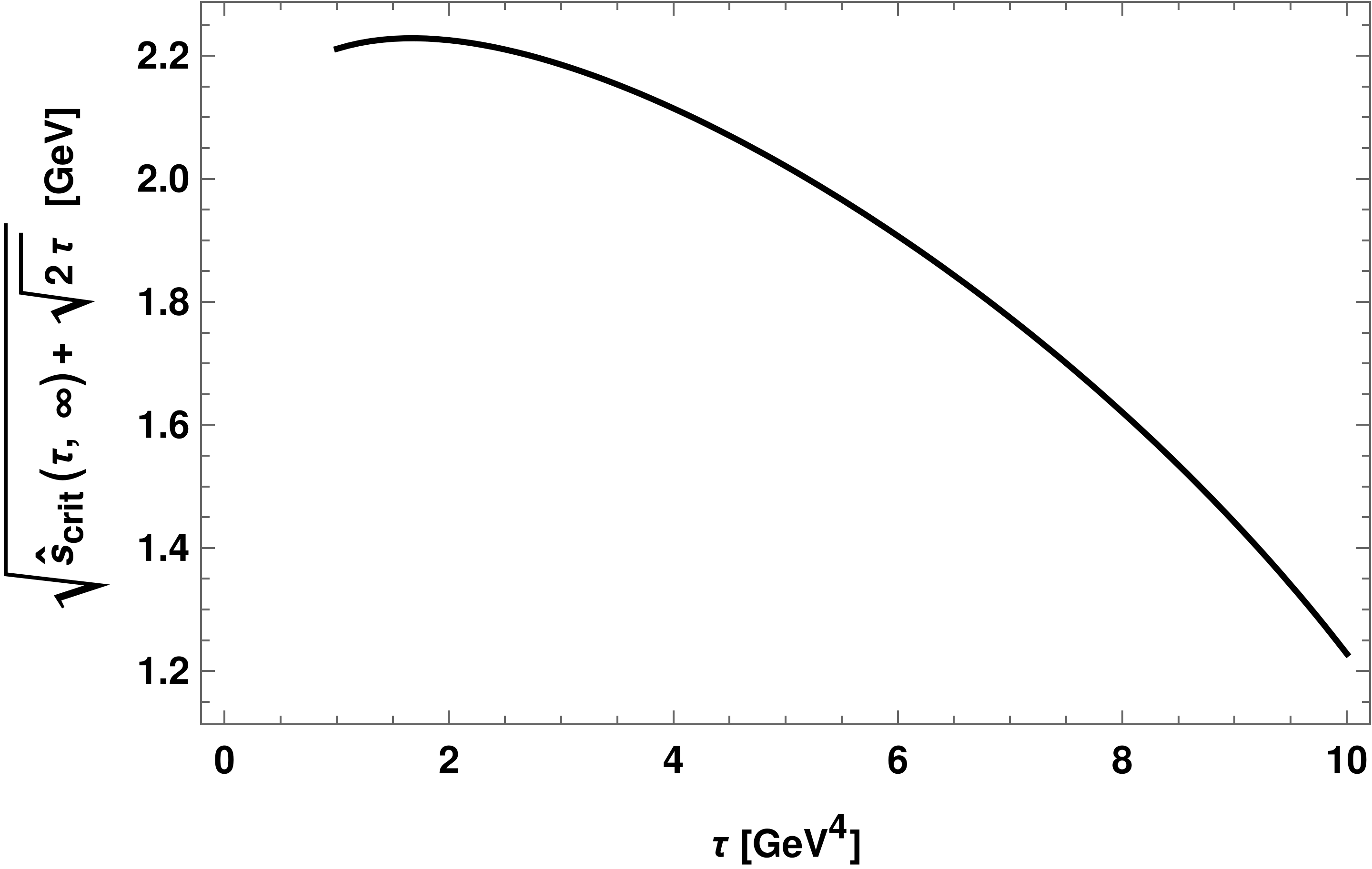}
\caption{The square-root of the RHS of~(\ref{ineq_inf}) versus $\tau$.}
\label{ineq_inf_Plot}
\end{figure}

Using the GSR, we motivate a lower bound on the \tq\ tetraquark ground-state mass $M$.
We employ a single narrow resonance model in~(\ref{res_plus_cont}),
\begin{equation}\label{snr_model}
    \rho^{\text{(had)}}(t) = f^2 \delta(t - M^2)
\end{equation}
where $M$ is the mass of the lightest $\tq$ tetraquark that couples to~(\ref{j1})
and $f$ is its corresponding coupling strength.
In using~(\ref{snr_model}), it is assumed that excited states are sufficiently 
suppressed relative to the ground state by the Gaussian kernel of~(\ref{gsr_defn})
that they can be ignored.
Plugging~(\ref{snr_model}) into~(\ref{sgsr}) gives
\begin{equation}\label{sgsr_snr}
     \gsr \triple{\hat{s}}{\tau}{s_0}
    = \frac{1}{\sqrt{4\pi\tau}}\int_{0}^{s_0}\! \ee^{-\frac{(t-\hat{s})^2}{4\tau}} 
      \rho^{\text{(OPE)}}(t)\,\dt
    =  \frac{f^2}{\sqrt{4\pi\tau}}\ee^{-\frac{(\hat{s}-M^2)^2}{4\tau}}.
\end{equation}
In general, in~(\ref{sgsr_snr}), $\tau$ should be restricted to some interval 
$\tau_{\min}\leq\tau\leq\tau_{\max}$.
For instance, $\tau_{\min}$ should be chosen such that the running coupling 
$\alpha_s(\sqrt[4]{\tau})$ (see~(\ref{running_alpha})) is not too large.
As in~\cite{Orlandini:2000nv}, we choose $\tau_{\min}=1~\gev^4$.
Regarding $\tau_{\max}$, the width of the Gaussian kernel on the RHS 
of~(\ref{sgsr}) is $\sqrt{2\tau}$, and,
as $\tau$ increases, so too does the sensitivity of the GSR
to excited states, eventually violating~(\ref{snr_model}).
Fortunately, for our purposes, we do not actually need a specific value for $\tau_{\max}$.
(For a more rigourous discussion of $\tau_{\min}$ and $\tau_{\max}$
based on H\"{o}lder inequalities, see~\cite{Ho:2018cat}.)
Since the RHS of~(\ref{sgsr_snr}) is positive,
it follows that $\gsr\triple{\hat{s}}{\tau}{s_0}$ must also be positive;
however, since $\rho^{\text{(OPE)}}(t)<0$ for $t\lesssim 4~\text{GeV}^2$, 
we find that there are regions of $\triple{\hat{s}}{\tau}{s_0}$ parameter space where 
$\gsr\triple{\hat{s}}{\tau}{s_0}<0$.
This is shown in Fig.~\ref{GSR_Plot} for $\tau=1~\text{GeV}^4$ and $s_0=10~\text{GeV}^2$.
Such regions of parameter space are unphysical.
We denote the zero of $\gsr\triple{\hat{s}}{\tau}{s_0}$ with respect to $\hat{s}$
as $\hat{s}_{\text{crit}}\double{\tau}{s_0}$.
We note that the single narrow resonance contribution to the RHS of~(\ref{sgsr_snr}) 
has width $\sqrt{2\tau}$. 
If, for self-consistency, we require that the full width of the resonance contribution
be contained in the region where $\gsr\triple{\hat{s}}{\tau}{s_0}>0$, then
\begin{equation}\label{ineq}
  M^2 \gtrsim \sqrt{2\tau} + \hat{s}_{\text{crit}}\double{\tau}{s_0}
\end{equation}
for all allowed $\tau$ at physical $s_0$.
Numerically, we find that the RHS of~(\ref{ineq}) is a decreasing function of $s_0$.
Thus,
\begin{equation}\label{ineq_inf}
  M^2 \gtrsim \sqrt{2\tau} + \hat{s}_{\text{crit}}\double{\tau}{\infty}.
\end{equation}
In Fig.~\ref{ineq_inf_Plot}, we plot the square root of the RHS of~(\ref{ineq_inf}) versus $\tau$
and find that $M \gtrsim 2.2$~\gev.
An analogous analysis that omits NLO perturbation theory finds that $M \gtrsim 2.4$~GeV.

\begin{figure}[htb]
\centering
\includegraphics[scale=0.8]{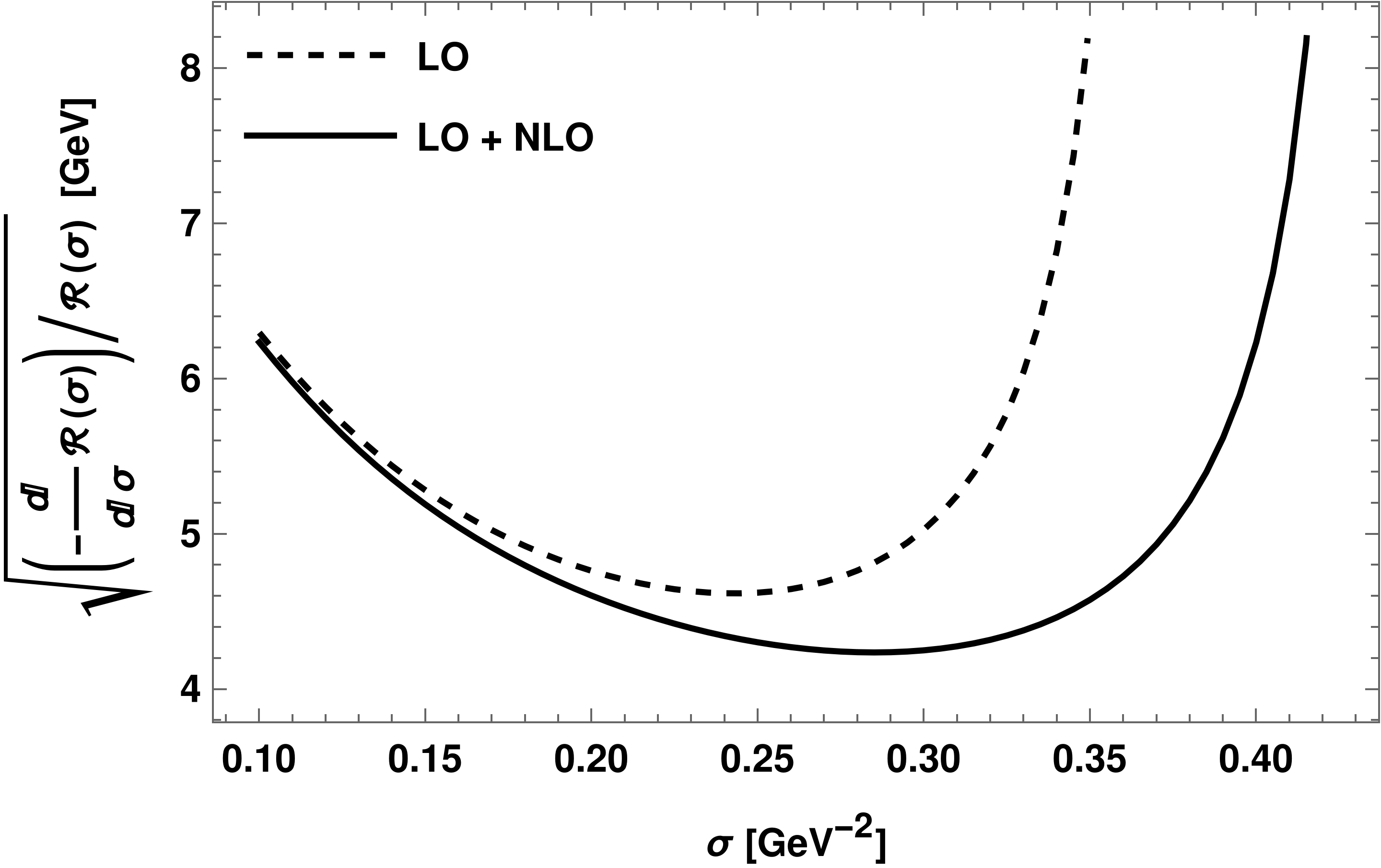}
\caption{$\sqrt{-\frac{\mathrm{d}\lsr(\sigma)}{\mathrm{d}\sigma}/\lsr(\sigma)}$ 
  with NLO perturbation theory (the solid line) and without (the dashed line).}
\label{lsr_ratio}
\end{figure}

Using the LSR, we determine an upper bound on the \tq\ tetraquark ground-state mass $M$.
As shown in~\cite{Harnett:2000xz},
\begin{equation}\label{ineq_lsr}
  M \leq \sqrt{\frac{-\frac{\mathrm{d}}{\mathrm{d}\sigma}\lsr(\sigma)}{\lsr(\sigma)}}.
\end{equation}
Inequality~(\ref{ineq_lsr}) follows from positivity of the hadronic spectral function
and applies to an extensive class of resonance models~\cite{Harnett:2000xz}.
In Fig.~\ref{lsr_ratio}, we plot the right-hand side of~(\ref{ineq_lsr}) 
with and without NLO perturbation theory.
Without NLO perturbation theory, we find that $M\leq4.6$~GeV.
With NLO perturbation theory, we find that $M\leq4.2$~GeV.
We note that, at the minimum value of the solid curve in Fig.~\ref{lsr_ratio}, 
the LSR $\lsr(\sigma)$ is positive and its perturbative contributions are greater than 
three times the (magnitude of the) condensate contributions,
\ie\ the extracted upper bound comes from a region of OPE convergence.

Uncertainty in our results is dominated by the value of $\kappa$  used in~(\ref{rho_cond}). 
For the central value $\kappa=2$ used in the above analysis, 
our NLO mass bounds are $2.2~\gev\lesssim M\leq4.2~\gev$.
Both upper and lower mass bounds increase with increasing $\kappa$.
For $\kappa=3$ (the upper range of Refs.~\cite{Narison:2002woh,Bertlmann:1987ty,Narison:1995jr}),
we find $2.4~\gev\lesssim M\leq4.6~\gev$. 
Because the prevailing evidence for violation of vacuum saturation indicates $\kappa>1$ \cite{Narison:2002woh,Bertlmann:1987ty,Narison:1995jr}, 
our $\kappa=1$ result of $1.9~\gev\lesssim M\leq4.0~\gev$ provides a conservative  
lower mass bound of $M>1.9~\gev$.

\section{Discussion}

Motivated by the unphysical violation of positivity in the LO 
LSRs of~\cite{Fu:2018ngx} 
and the large NLO perturbative effects for  $0^{++}$ light tetraquarks~\cite{Groote:2014pva}, we
calculated NLO contributions to perturbation theory for a \tq\ tetraquark correlation 
function~(\ref{correlator})--(\ref{Pi_zero}) in the limit of massless $u$ and $d$ quarks.  
Our results represent the first complete NLO perturbative calculation of light-quark, 
exotic-$J^{PC}$ tetraquark sum rules.
Instead of renormalizing the interpolating current~(\ref{j1}), 
we eliminated nonlocal divergences using diagrammatic renormalization methods as outlined in 
Ref.~\cite{deOliveira:2022eeq}.
Instead of evaluating dimensionally regularized integrals analytically, we evaluated 
them numerically using pySecDec~\cite{Borowka:2017idc}.
We then fit the pySecDec-generated data to the known functional form of 
the imaginary part of the correlation function.
The successful combination of pySecDec with
diagrammatic renormalization establishes a valuable and efficient new methodology for computing radiative 
corrections to correlation functions of operators composed of light quarks.
Furthermore, diagrammatic renormalization and pySecDec can, in principle,
be applied to systems containing heavy quarks
although, in our experience, computational runtimes and RAM requirements increase
significantly when heavy quarks are introduced.

Relative to LO perturbation theory, the NLO corrections make significant contributions
to $\rho^{\text{(OPE)}}(t)$ (see Figs.~\ref{rho_ratio}--\ref{rho_ope}) 
and to QCD sum rules (see Figs.~\ref{lo_nlo_lsrs_plot}--\ref{GSR_Plot}).
Although the NLO corrections mitigate the violation of positivity in the sum rules, 
there are still unphysical regions of $s_0$ and Borel-scale parameter space.
Using positivity to constrain physical regions within a GSR, 
we motivated a lower bound on the $\tq$ tetraquark ground-state mass $M$, and,
using an LSR, we determined an upper bound.
Taking into account both LO and NLO perturbation theory, we found that,
for our central vacuum saturation parameter $\kappa=2$,
\begin{equation}
  2.2~\text{GeV} \lesssim M \leq 4.2~\text{GeV}
  \label{final_bounds}
\end{equation}
which should be compared with a range determined by omitting NLO perturbation theory,
\begin{equation}
  2.4~\text{GeV} \lesssim M \leq 4.6~\text{GeV}.
\end{equation}
Increasing the vacuum saturation parameter to $\kappa=3$ increases both the upper 
and lower bounds in~\eqref{final_bounds}.
For the current~(\ref{j1}), we found no evidence for the existence of a 
$\tq$ tetraquark under 1.9~GeV, and note that 1.9~\gev\ was
obtained by decreasing the vacuum saturation parameter to $\kappa=1$, 
thereby underestimating known violations of vacuum saturation~\cite{Narison:2002woh,Bertlmann:1987ty,Narison:1995jr}.
This lower bound on the $\tq$ mass is contrary to the results of~\cite{Fu:2018ngx}.
The source of the discrepancy between Ref.~\cite{Fu:2018ngx} 
and our conservative mass bound $M>1.9~\gev$ can be traced to 
the value of $s_0$ used in~\cite{Fu:2018ngx} which
violates the physical positivity constraint~(\ref{s0_bound_lo}).
Finally, we note that the \tq\ mass bounds~\eqref{final_bounds} encompass the 
QCD Gaussian sum-rules mass predictions 
for $0^{+-}$, light-quark hybrids~\cite{Ho:2018cat}, 
suggesting the interesting possibility of hybrid-tetraquark mixing in light-quark systems.

\section*{Acknowledgments}
The work is supported by the Natural Sciences and Engineering Research Council of Canada (NSERC).
We are grateful to Wei Chen and Zhou-Ran Huang for valuable discussions.

\end{document}